\documentclass[journal]{IEEEtran}
\pdfoutput=1
\usepackage[T1]{fontenc}
\usepackage[latin9]{inputenc}
\usepackage{amsmath}
\usepackage{amsthm}
\usepackage{amssymb}
\usepackage{stackrel}
\usepackage{graphicx}
\usepackage[unicode=true,
 bookmarks=true,bookmarksnumbered=true,bookmarksopen=true,bookmarksopenlevel=1,
 breaklinks=false,pdfborder={0 0 0},pdfborderstyle={},backref=false,colorlinks=false]
 {hyperref}
\hypersetup{pdftitle={Your Title},
 pdfauthor={Your Name},
 pdfpagelayout=OneColumn, pdfnewwindow=true, pdfstartview=XYZ, plainpages=false}

\makeatletter

\theoremstyle{plain}
\newtheorem{thm}{\protect\theoremname}
\theoremstyle{plain}
\newtheorem{prop}[thm]{\protect\propositionname}

\usepackage[caption=false,font=footnotesize]{subfig}
\usepackage{cite}

\makeatother

\providecommand{\propositionname}{Proposition}
\providecommand{\theoremname}{Theorem}

\begin{document}
\title{Energy-Efficient Hybrid Beamforming with Dynamic On-off Control for
Integrated Sensing, Communications, and Powering}
\author{Zeyu~Hao, Yuan~Fang,~\IEEEmembership{Member,~IEEE,} Xianghao~Yu,~\IEEEmembership{Member,~IEEE,}
Jie~Xu,~\IEEEmembership{Senior~Member,~IEEE,} Ling~Qiu,~\IEEEmembership{Member,~IEEE,}
Lexi~Xu,~\IEEEmembership{Senior~Member,~IEEE,} and~Shuguang~Cui,~\IEEEmembership{Fellow,~IEEE}\thanks{Z.~Hao and L.~Qiu are with the Key Laboratory of Wireless-Optical
Communications, Chinese Academy of Sciences, School of Information
Science and Technology, University of Science and Technology of China,
Hefei 230027, China (e-mail: \protect\href{mailto:hzy2018@mail.ustc.edu.cn}{hzy2018@mail.ustc.edu.cn};
\protect\href{mailto:lqiu@ustc.edu.cn}{lqiu@ustc.edu.cn}).}\thanks{Y.~Fang is with the Future Network of Intelligence Institute (FNii),
The Chinese University of Hong Kong (Shenzhen), Shenzhen 518172, China,
and the Department of Electrical Engineering, City University of Hong
Kong (CityU), Hong Kong (e-mail: \protect\href{mailto:yuanfang@cityu.edu.hk}{yuanfang@cityu.edu.hk}).}\thanks{X.~Yu is with the Department of Electrical Engineering, City University
of Hong Kong (CityU), Hong Kong (e-mail: \protect\href{mailto:alex.yu@cityu.edu.hk}{alex.yu@cityu.edu.hk}).}\thanks{J.~Xu, and S.~Cui are with the School of Science and Engineering
(SSE), and FNii, The Chinese University of Hong Kong (Shenzhen), Shenzhen
518172, China (e-mail: \protect\href{mailto:xujie@cuhk.edu.cn}{xujie@cuhk.edu.cn};
\protect\href{mailto:shuguangcui@cuhk.edu.cn}{shuguangcui@cuhk.edu.cn}).}\thanks{L.~Xu is with the Research Institute, China United Network Communications
Corporation, Beijing 100048, China (e-mail: \protect\href{mailto:davidlexi@hotmail.com}{davidlexi@hotmail.com}).}\thanks{L.~Qiu and J.~Xu are the corresponding authors. }}
\maketitle
\begin{abstract}
This paper investigates the energy-efficient hybrid beamforming design
for a multi-functional integrated sensing, communications, and powering
(ISCAP) system. In this system, a base station (BS) with a hybrid
analog-digital (HAD) architecture sends unified wireless signals to
communicate with multiple information receivers (IRs), sense multiple
point targets, and wirelessly charge multiple energy receivers (ERs)
at the same time. To facilitate the energy-efficient design, we present
a novel HAD architecture for the BS transmitter, which allows dynamic
on-off control of its radio frequency (RF) chains and analog phase
shifters (PSs) through a switch network. We also consider a practical
and comprehensive power consumption model for the BS, by taking into
account the power-dependent non-linear power amplifier (PA) efficiency,
and the on-off non-transmission power consumption model of RF chains
and PSs. We jointly design the hybrid beamforming and dynamic on-off
control at the BS, aiming to minimize its total power consumption,
while guaranteeing the performance requirements on communication rates,
sensing Cram\'er-Rao bound (CRB), and harvested power levels.
The formulation also takes into consideration the per-antenna transmit
power constraint and the constant modulus constraints for the analog
beamformer at the BS. The resulting optimization problem for ISCAP
is highly non-convex due to the binary on-off non-transmission power
consumption of RF chains and PSs, the non-linear PA efficiency, and
the coupling between analog and digital beamformers. To tackle this
problem, we first approximate the binary on-off non-transmission power
consumption into a continuous form, and accordingly propose an iterative
algorithm to find a high-quality approximate solution with ensured
convergence, by employing techniques from alternating optimization
(AO), sequential convex approximation (SCA), and semi-definite relaxation
(SDR). Then, based on the optimized beamforming weights, we develop
an efficient method to determine the binary on-off control of RF chains
and PSs, as well as the associated hybrid beamforming solution. Numerical
results show that the proposed design achieves an improved energy
efficiency for ISCAP than other benchmark schemes without joint design
of hybrid beamforming and dynamic on-off control. This validates the
benefit of dynamic on-off control in energy reduction, especially
when the multi-functional performance requirements become less stringent.
\end{abstract}

\begin{IEEEkeywords}
Integrated sensing, communication, and powering (ISCAP), energy efficiency,
hybrid beamforming, dynamic on-off control, non-linear power amplifier
(PA) efficiency.
\end{IEEEkeywords}

\section{Introduction}

With recent advancements in artificial intelligence (AI) of things
(AIoT), future sixth-generation (6G) wireless networks are envisioned
to incorporate a large number of AIoT devices, which are equipped
with sophisticated sensing, communication, computation, and control
capabilities to perform various AI tasks. Towards this end, wireless
networks are experiencing a paradigm shift from conventional communication-only
systems in fifth-generation (5G) to new multi-functional systems integrating
communication, sensing, computation, control, and even wireless power
transfer (WPT) in 6G, thus supporting the sustainable operation of
AIoT devices \cite{Zhu2023}. Recently, integrated sensing and communication
(ISAC) and wireless information and power transfer (WIPT) have attracted
extensive research interests from both academia and industry \cite{9737357,8476597},
in which radio signals in wireless communication networks are reused
for dual functions of environmental sensing and wireless energy delivery
for charging low-power energy receivers (ERs), respectively. As ISAC
and WIPT technologies mature and spectrum resources become increasingly
constrained, it naturally leads to their combined integration at the
air interface for 6G. This thus introduces new triple-functional wireless
systems with integrated sensing, communication, and powering (ISCAP).
Such triple-functional integration is expected to not only enhance
the resource utilization efficiency and reduce system cost, but also
foster mutual benefits among different functionalities \cite{2024arXiv240103516C}.

There have been a handful of works in the recent literature considering
the ISCAP system \cite{10382465,2023arXiv231109028Z,2023arXiv231100104Z}.
The work \cite{10382465} first exploited the multiple-input multiple-output
(MIMO) technology for ISCAP systems. It focused on optimizing the
transmit signal covariance at the multi-functional BS to reveal the
fundamental performance tradeoff among three functionalities, in terms
of sensing Cram\'er-Rao bound (CRB), communication rate,
and harvested energy, respectively, by characterizing the Pareto boundary
of the so-called CRB-rate-energy (C-R-E) region. The authors in \cite{2023arXiv231109028Z}
then studied the transmit beamforming design for a multi-user multi-antenna
ISCAP system, with the objective of optimizing the sensing performance
while ensuring communication and WPT requirements. Furthermore, the
orthogonal frequency division multiplexing (OFDM) technique was employed
for ISCAP in \cite{2023arXiv231100104Z} by considering the non-zero
mean asymmetric Gaussian distributed signaling for each subcarrier,
in which the mean and variance at subcarriers were designed to maximize
the harvested power for WPT while ensuring the requirements on both
the communications rate and the average side-to-peak-lobe difference
for sensing. In general, these prior works focused on spectral-efficient
ISCAP designs, with the objective of maximizing the multi-functional
performance subject to the constraints on spectrum and power resources.

The multi-functional integration in ISCAP, however, results in ever-increasing
service requirements. First, sensing requirements strongly depend
on the line-of-sight (LoS) links between transceivers and sensing
targets, and the transmission range for WPT is generally much shorter
than that for wireless communications. Therefore, more base station
(BS) infrastructures with massive antennas need to be densely deployed
to ensure adequate network coverage for supporting multi-functional
services. However, this unavoidably leads to increased operational
expenses and network energy consumption. Second, the deployment of
BS infrastructures is planned to support the peak traffic loads, but
the traffic loads may fluctuate significantly over both time and space.
This consequently can result in energy wastage during off-peak hours,
owing to the non-transmission energy consumption caused by radio frequency
(RF) chains, digital-to-analog converters (DACs), etc. \cite{10328645}.
Consequently, conventional spectral-efficient ISCAP designs may not
work well in energy-limited scenarios, resulting in heightened energy
consumption and increased carbon emissions. Therefore, the investigation
of energy-efficient ISCAP design is both significant and pressing,
which motivates the current work. To our best knowledge, although
several existing works have focused on investigating the energy efficiency
of ISAC \cite{9761984,10445319,2023arXiv231017401Z} and WIPT \cite{9133120,8233108,8478252}
systems under certain power and performance constraints, the energy-efficient
design for ISCAP systems remains inadequately understood.

In order to reduce hardware complexity and power consumption, hybrid
analog-digital (HAD) architecture has been widely considered as an
enabling technique in MIMO wireless systems \cite{2022arXiv220801235E,8254864}.
With HAD architecture at the BS, a restricted number of RF chains
are linked to an extensive array of antennas via a network of analog
components such as switches and/or phase shifters (PSs) \cite{6717211}.
Prior works have developed various HAD architectures to balance the
tradeoff between system cost and performance. For instance, the fully-connected
HAD architecture \cite{9729809} allows each RF chain to be connected
to all antennas, thus offering more flexibility in beamforming optimization.
By contrast, the partially-connected HAD architecture offers reduced
hardware complexity and lower power consumption while sacrificing
the flexibility in beamforming design and compromising the performance
\cite{9868348}. In order to enhance the energy efficiency of HAD
design, the authors in \cite{9094668} proposed to switch off the
RF chain to transmit antenna connections for saving the energy consumption
of the associated components. The work \cite{8382230} analyzed the
energy efficiency of different HAD architectures, which shows that
dynamically selecting PSs via properly connected switches contributes
to a substantial enhancement in energy efficiency without sacrificing
spectrual efficiency in general. However, these prior works on energy-efficient
HAD design only considered the transmit power consumption, while ignoring
various practical issues like the power-dependent non-linear power
amplifier (PA) efficiency \cite{7264986,9133207,10304516} and on-off
non-transmission power consumption of RF chains \cite{10328645,10304516}
and PSs. In fact, it is widely known that the consideration of on-off
non-transmission power has a significant impact on the energy efficiency
of wireless communications \cite{5783982}, and the dynamic on-off
control of the corresponding components including RF chains and PSs
is essential in enhancing the energy efficiency. It is also established
in \cite{10304516} that the consideration of non-linear PA efficiency
can affect the dynamic on-off behaviors and thus is important for
further improving the energy efficiency performance. Therefore, by
considering such practical and comprehensive power consumption models,
how to design the HAD architecture and optimize the energy efficiency
is an interesting but unaddressed problem, especially for the ISCAP
system.

Building upon the aforementioned unresolved issues, this paper investigates
the energy-efficient design for ISCAP systems with an HAD multi-functional
BS serving multiple information receivers (IRs) and ERs, and sensing
multiple targets at the same time. The main results are listed as
follows. 
\begin{itemize}
\item First, to facilitate the energy-efficient ISCAP design, we consider
a novel HAD architecture, in which a switch network is employed to
enable the dynamic on-off control of RF chains and PSs for energy
saving. Under this setup, we consider a comprehensive power consumption
model for the BS by taking into account the practical non-linear PA
efficiency and on-off non-transmission power consumption of both RF
chains and PSs. 
\item Next, we present the joint information and sensing/energy beamforming
design, by allowing the BS to send dedicated sensing/energy signals
together with information signals. Accordingly, we optimize the digital
beamforming for joint information/sensing/energy transmission, together
with the analog beamforming as well as the dynamic on-off switching
of RF chains and PSs for enhancing the energy efficiency. Our objective
is to minimize the total power consumption at the BS subject to constraints
on communication, sensing, and powering performances, as well as those
on per-antenna transmit power and constant modulus for analog beamforming.
The formulated problem exhibits high non-convexity due to the binary
on-off non-transmission power consumption of RF chains and PSs, the
non-linear PA efficiency, the element-wise constant modulus constraint
for analog beamforming, and the coupling between analog and digital
beamformers.
\item Furthermore, to tackle this optimization problem, we first approximate
the binary on-off on-transmission power consumption into a continuous
form, and then optimize the analog and digital beamformers alternately
based on alternating optimization (AO), in which the semi-definite
relaxation (SDR) and sequential convex approximation (SCA) techniques
are employed to deal with the non-convexity caused by the beamformers
and non-linear PA efficiency. Based on the obtained hybrid beamforming
weights, we propose an efficient method to determine the on-off control
of RF chains and PSs, and accordingly decide the hybrid beamforming
solution.
\item Finally, numerical results show that the proposed design exhibits
superior power-saving performance for ISCAP compared to benchmark
schemes without joint hybrid beamforming and dynamic on-off control.
In particular, the dynamic on-off control is shown to be particularly
beneficial when the multi-functional performance requirements become
less strict. It is also shown that under the non-linear PA efficiency
model, the BS tends to switch off more antennas and the associated
PSs/RF chains for saving power.
\end{itemize}
\quad{}The remainder of this paper is organized as follows. Section
II introduces the ISCAP system model with HAD architecture at BS,
and formulates the problem. Section III proposes an efficient algorithm
to solve the highly non-convex optimization problem. Section IV presents
numerical results. Section V concludes the paper.

\textit{Notations}: We use boldface letters to represent vectors (lower-case)
and matrices (upper-case). For a complex-valued element $a$, $\Re\left(a\right)$
and $\Im\left(a\right)$ denote its real and imaginary parts, respectively,
and $\left|a\right|$ denotes its modulus. For a complex vector $\boldsymbol{a}$,
$\left\Vert \boldsymbol{a}\right\Vert $ denotes its Euclidean norm,
and $\textrm{diag}\left(\boldsymbol{a}\right)$ denotes a diagonal
matrix with $\boldsymbol{a}$ being its diagonal entries. For a square
matrix $\boldsymbol{A}$, $\textrm{tr}(\boldsymbol{A})$ denotes its
trace and $\boldsymbol{A}\succeq\boldsymbol{0}$ indicates that $\boldsymbol{A}$
is positive semi-definite. For a complex matrix $\boldsymbol{B}$
of arbitrary size, we use $\mathrm{rank}\left(\boldsymbol{B}\right)$,
$\boldsymbol{B}^{c}$, $\boldsymbol{B}^{T}$, and $\boldsymbol{B}^{H}$
to denote its rank, complex conjugate, transpose, and conjugate transpose,
respectively, and use $\left[\boldsymbol{B}\right]_{i,j}$ to denote
its $\left(i,j\right)$-th element and $\textrm{vec}\left(\boldsymbol{B}\right)$
to denote its column vectorization. Moreover, $\mathbb{E}\left[\cdot\right]$
denotes the statistical expectation, and $\odot$ denotes the Hadamard
product. $\boldsymbol{I}_{x}$ denotes the identity matrix with the
size of $x\times x$. $\boldsymbol{1}_{n}^{m}\in\mathbb{\mathbb{R}}^{1\times m}$
is a vector with its $n$-th element being one and others zero, and
$\boldsymbol{E}_{n}^{m}\in\mathbb{C}^{m\times m}$ is a square matrix
with the $n$-th diagonal element being 1 and the others being zero.
The indicator function $\mathcal{I}\left\{ \cdot\right\} $ is defined
as $\mathcal{I}\left\{ x\right\} =0$ if $x=0$, and $\mathcal{I}\left\{ x\right\} =1$
otherwise.

\section{System Model and Problem Formulation}

\begin{figure}[tbh]
\begin{centering}
\includegraphics[width=3.5in]{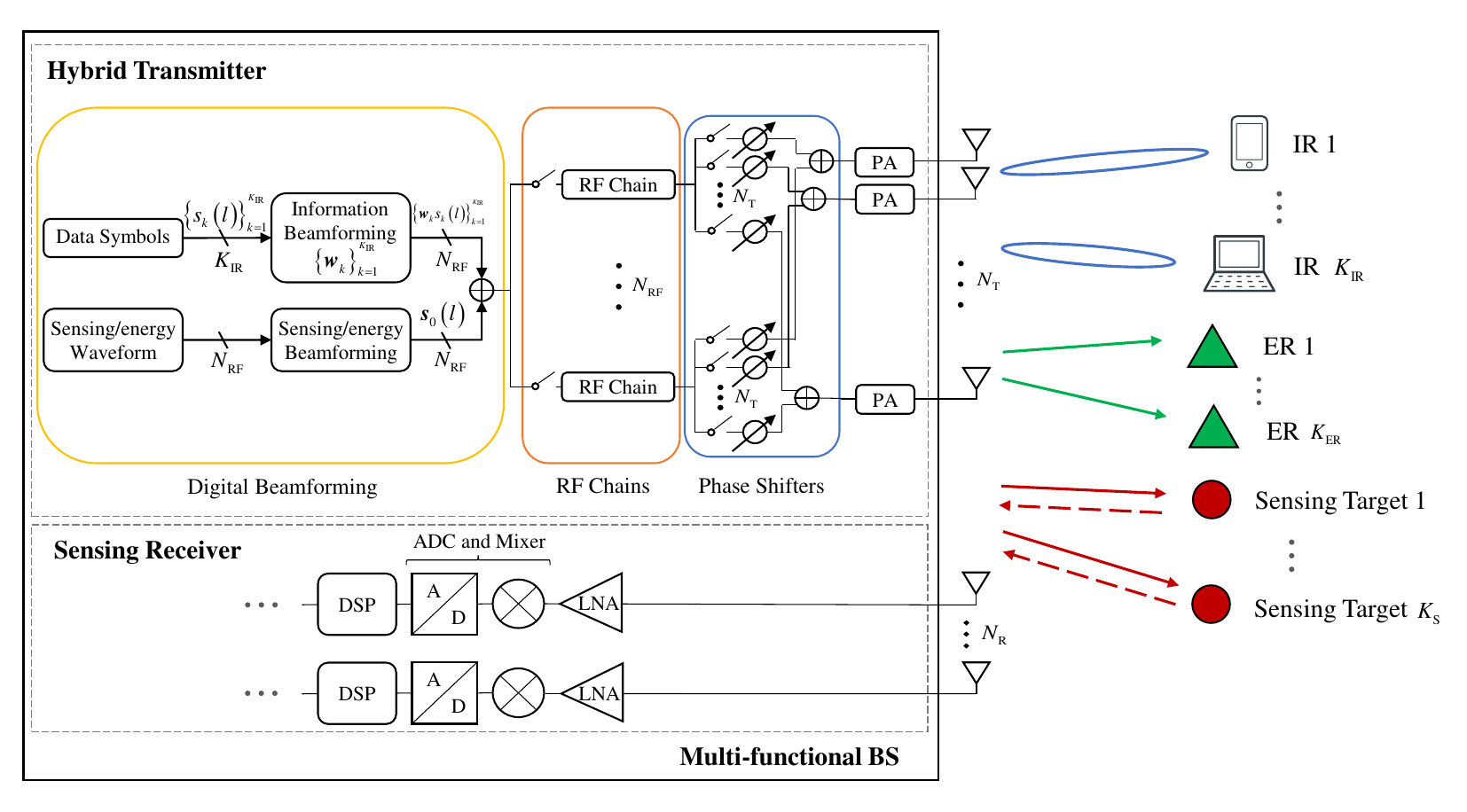}
\par\end{centering}
\caption{The ISCAP system with an HAD multi-functional BS.\label{Fig. 1}}
\end{figure}

We consider a downlink multi-antenna ISCAP system as shown in Fig. 1,
where a BS communicates with $K_{\textrm{IR}}$ single-antenna IRs,
wirelessly charges $K_{\textrm{ER}}$ single-antenna ERs, and senses
$K_{\textrm{S}}$ point targets at the same time. Let $\mathcal{K}_{\textrm{IR}}\triangleq\left\{ 1,\ldots,K_{\textrm{IR}}\right\} $,
$\mathcal{K}_{\textrm{ER}}\triangleq\left\{ 1,\ldots,K_{\textrm{ER}}\right\} $,
and $\mathcal{K}_{\textrm{S}}\triangleq\left\{ 1,\ldots,K_{\textrm{S}}\right\} $
denote the sets of IRs, ERs, and sensing targets, respectively. The
multi-functional BS consists of an HAD transmitter and a fully-digital
sensing receiver. We assume that the HAD transmitter is equipped with
$N_{\textrm{T}}$ uniform linear array (ULA) antennas, $N_{\textrm{RF}}$
RF chains, and $N_{\textrm{T}}N_{\textrm{RF}}$ PSs, while the sensing
receiver has $N_{\textrm{R}}$ ULA antennas. Furthermore, each RF
chain and PS is connected to a switch, enabling their dynamic on-off
control to facilitate energy-efficient operation. Let $\mathcal{N}_{\textrm{T}}\triangleq\{1,\ldots,N_{\textrm{T}}\}$
and $\mathcal{N}_{\textrm{RF}}\triangleq\{1,\ldots,N_{\textrm{RF}}\}$
denote the sets of transmit antennas and RF chains at the transmitter,
respectively. It is assumed that $K_{\textrm{IR}}\leq N_{\textrm{RF}}\leq N_{\textrm{T}}\leq N_{\textrm{R}}$
to ensure the spatial degrees of freedom (DoFs) for communication
and sensing.

We consider a quasi-static narrowband channel model, assuming that
the wireless channels remain constant throughout the transmission
block of interest. Let $\boldsymbol{h}_{k}^{H}\in\mathbb{C}^{1\times N_{\textrm{T}}}$
and $\boldsymbol{d}_{j}^{H}\in\mathbb{C}^{1\times N_{\textrm{T}}}$
denote the channel vectors from the BS to IR $k$ and ER $j$, respectively,
which are assumed to be known at the BS via proper channel estimation
\cite{9440479}. Let $\boldsymbol{G}_{i}\in\mathbb{C}^{N_{\textrm{R}}\times N_{\textrm{T}}}$
denote the target response matrix from the BS transmitter to target
$i$ to the BS sensing receiver.

We consider a joint transmit beamforming design for ISCAP, in which
dedicated sensing/energy beams are employed together with the information
beams for multi-functional transmission\footnote{The dedicated sensing/energy signals are considered to provide sufficient
DoFs for sensing and WPT, which is particularly useful when the number
of IRs $K_{\textrm{IR}}$ becomes small or zero. }. Let $s_{k}\left(l\right)$ denote the information signal for IR
$k$ at symbol $l$ with $\mathbb{E}\left[\left|s_{k}\left(l\right)\right|^{2}\right]=1$.
Let $\boldsymbol{s}_{\textrm{0}}\left(l\right)\in\mathbb{C}^{N_{\textrm{RF}}\times1}$
denote the dedicated sensing/energy signal, which is independent from
$\left\{ s_{k}\left(l\right)\right\} $ and with covariance matrix
$\boldsymbol{S}=\mathbb{E}\left[\boldsymbol{s}_{\textrm{0}}\left(l\right)\boldsymbol{s}_{\textrm{0}}^{H}\left(l\right)\right]\succeq0$.
Let $\boldsymbol{w}_{k}\in\mathbb{C}^{N_{\textrm{RF}}\times1}$ denote
the transmit digital beamformer for IR $k\in\mathcal{K}_{\textrm{IR}}$,
and $\boldsymbol{F}\in\mathbb{C}^{N_{\textrm{T}}\times N_{\textrm{RF}}}$
denote the analog beamformer. Combing information and sensing/energy
signals, the transmitted baseband signal by the BS is

\begin{equation}
\boldsymbol{x}\left(l\right)=\underset{k\in\mathcal{K}_{\textrm{IR}}}{\sum}\boldsymbol{F}\boldsymbol{w}_{k}s_{k}\left(l\right)+\boldsymbol{F}\boldsymbol{s}_{\textrm{0}}\left(l\right),\label{eq:x(L)}
\end{equation}
where the covariance matrix of $\boldsymbol{x}\left(l\right)$ is
\begin{equation}
\boldsymbol{R}_{\boldsymbol{x}}=\mathbb{E}\left(\boldsymbol{x}\left(l\right)\boldsymbol{x}^{H}\left(l\right)\right)=\underset{k\in\mathcal{K}_{\textrm{IR}}}{\sum}\boldsymbol{F}\boldsymbol{w}_{k}\boldsymbol{w}_{k}^{H}\boldsymbol{F}^{H}+\boldsymbol{F}\boldsymbol{S}\boldsymbol{F}^{H}.\label{eq:RX}
\end{equation}

Note that we consider the dynamic on-off control of PSs and RF chains.
Therefore, it follows that $\left[\boldsymbol{F}\right]_{i,j}\in\mathcal{F}\triangleq\left\{ x\mid\left|x\right|\in\left\{ 0,\frac{1}{\sqrt{N_{\textrm{T}}}}\right\} \right\} $,
where $\left|\left[\boldsymbol{F}\right]_{i,j}\right|=\frac{1}{\sqrt{N_{\textrm{T}}}}$
when the corresponding PS element is activated, and $\left[\boldsymbol{F}\right]_{i,j}=0$
when the PS element is turned off to be inactive. Similarly, for each
RF chain $n\in\mathcal{N}_{\textrm{RF}}$, when it is activated, we
have $\sum_{k\in\mathcal{K}_{\textrm{IR}}}\left|\boldsymbol{1}_{n}^{N_{\textrm{RF}}}\boldsymbol{w}_{k}\right|^{2}+\textrm{tr}\left(\boldsymbol{E}_{n}^{N_{\textrm{RF}}}\boldsymbol{S}\right)>0$,
denoting that the digital signal stream from the digital beamforming
is non-zero; while when it is turned off, it holds that $\sum_{k\in\mathcal{K}_{\textrm{IR}}}\left|\boldsymbol{1}_{n}^{N_{\textrm{RF}}}\boldsymbol{w}_{k}\right|^{2}+\textrm{tr}\left(\boldsymbol{E}_{n}^{N_{\textrm{RF}}}\boldsymbol{S}\right)=0$,
denoting that the digital signal stream is zero.

\subsection{Communication Model}

First, we consider the downlink multi-user communication from the
BS to IRs. According to the transmit signal in \eqref{eq:x(L)}, the
received signal at IR $k$ is expressed as
\begin{align}
y_{k}\left(l\right) & =\boldsymbol{h}_{k}^{H}\boldsymbol{x}\left(l\right)+z_{k}\left(l\right)\nonumber \\
 & =\underset{\textrm{desired\,signal}}{\underbrace{\boldsymbol{h}_{k}^{H}\boldsymbol{F}\boldsymbol{w}_{k}s_{k}\left(l\right)}}+\underset{\textrm{inter-IR\,interference}}{\underbrace{\boldsymbol{h}_{k}^{H}\underset{i\in\mathcal{K}_{\textrm{IR}},i\neq k}{\sum}\boldsymbol{F}\boldsymbol{w}_{i}s_{i}\left(l\right)}}\\
 & +\underset{\textrm{interference\,from\,sensing/energy\,signals}}{\underbrace{\boldsymbol{h}_{k}^{H}\boldsymbol{F}\boldsymbol{s}_{\textrm{0}}\left(l\right)}}+z_{k}\left(l\right),
\end{align}
where $z_{k}\left(l\right)\sim\mathcal{CN}\left(0,\sigma_{k}^{2}\right)$
denotes the additive white Gaussian noise (AWGN) at IR $k$, with
$\sigma_{k}^{2}$ denoting the noise power. As a result, the SINR
at IR $k\in\mathcal{K}_{\textrm{IR}}$ is given by

\begin{align}
 & \gamma_{k}\left(\boldsymbol{F},\left\{ \boldsymbol{w}_{k}\right\} ,\boldsymbol{S}\right)\nonumber \\
 & =\frac{\left|\boldsymbol{h}_{k}^{H}\boldsymbol{F}\boldsymbol{w}_{k}\right|^{2}}{\underset{i\in\mathcal{K}_{\textrm{IR}},i\neq k}{\sum}\left|\boldsymbol{h}_{k}^{H}\boldsymbol{F}\boldsymbol{w}_{i}\right|^{2}+\left|\boldsymbol{h}_{k}^{H}\boldsymbol{F}\boldsymbol{S}\boldsymbol{F}^{H}\boldsymbol{h}_{k}\right|+\sigma_{k}^{2}}.
\end{align}

\subsection{Sensing Model}

Next, we consider the monostatic radar sensing with multiple point
targets. In this case, the corresponding target response matrix is
given by

\begin{equation}
\boldsymbol{G}_{i}=\beta_{i}\boldsymbol{a}\left(\theta_{i}\right)\boldsymbol{v}^{T}\left(\theta_{i}\right).\label{eq:Gi}
\end{equation}
In \eqref{eq:Gi}, $\beta_{i}$ denotes the complex coefficient of
target $i$, depending on both the round-trip path-loss and the radar-cross-sections
(RCS). Furthermore, $\boldsymbol{v}\left(\theta_{i}\right)\in\mathbb{C}^{N_{\textrm{T}}\times1}$
and $\boldsymbol{a}\left(\theta_{i}\right)\in\mathbb{C}^{N_{\textrm{R}}\times1}$
denote the steering vectors of the transmit and receive antennas,
respectively, with $\theta_{i}$ denoting the direction of arrival
(DoA) of target $i$. Assuming half-wavelength spacing between adjacent
antennas and choosing the center of the ULA antennas as the reference
point, we have{\small{}
\begin{align}
\boldsymbol{v}\left(\theta_{i}\right) & =\left[e^{-j\frac{N_{\textrm{T}}-1}{2}\pi\sin\theta_{i}},e^{-j\frac{N_{\textrm{T}}-3}{2}\pi\sin\theta_{i}},\ldots,e^{j\frac{N_{\textrm{T}}-1}{2}\pi\sin\theta_{i}}\right]^{T},\nonumber \\
\boldsymbol{a}\left(\theta_{i}\right) & =\left[e^{-j\frac{N_{\textrm{R}}-1}{2}\pi\sin\theta_{i}},e^{-j\frac{N_{\textrm{R}}-3}{2}\pi\sin\theta_{i}},\ldots,e^{j\frac{N_{\textrm{R}}-1}{2}\pi\sin\theta_{i}}\right]^{T}.\label{eq:Steering_Vectors}
\end{align}
}The partial derivatives of $\boldsymbol{v}\left(\theta_{i}\right)$
and $\boldsymbol{a}\left(\theta_{i}\right)$ with respect to (w.r.t.)
$\theta_{i}$ are respectively given by{\small{}
\begin{align}
\dot{\boldsymbol{v}}\left(\theta_{i}\right) & =\left[-jv_{1}\frac{N_{\textrm{T}}-1}{2}\pi\cos\theta_{i},\ldots,jv_{N_{\textrm{T}}}\frac{N_{\textrm{T}}-1}{2}\pi\cos\theta_{i}\right]^{T},\nonumber \\
\dot{\boldsymbol{a}}\left(\theta_{i}\right) & =\left[-ja_{1}\frac{N_{\textrm{R}}-1}{2}\pi\cos\theta_{i},\ldots,ja_{N_{\textrm{R}}}\frac{N_{\textrm{R}}-1}{2}\pi\cos\theta_{i}\right]^{T},
\end{align}
}where $v_{i}$ and $a_{i}$ denote the $i$-th entries of $\boldsymbol{v}\left(\theta_{i}\right)$
and $\boldsymbol{a}\left(\theta_{i}\right)$, respectively.

The information and sensing/energy signals are jointly exploited to
sense the targets. Consequently, the received echo signal by the sensing
receiver at the BS is given by
\begin{equation}
\boldsymbol{y}_{\textrm{S}}\left(l\right)=\underset{i\in\mathcal{K}_{\textrm{S}}}{\sum}\boldsymbol{G}_{i}\boldsymbol{x}\left(l\right)+\boldsymbol{z}_{\textrm{S}}\left(l\right),\label{eq:Ys}
\end{equation}
where $\boldsymbol{z}_{\textrm{S}}\left(l\right)\sim\mathcal{CN}\left(\boldsymbol{0},\sigma_{\textrm{S}}^{2}\boldsymbol{I}_{N_{\textrm{R}}}\right)$
denotes the noise at the sensing receiver, with $\sigma_{\textrm{S}}^{2}$
denoting the noise power. Let $\mathcal{L}\triangleq\left\{ 1,\ldots,L\right\} $
denote the set of time symbols with $L$ being the radar dwell time.
The received echo signal during the radar dwell time is expressed
as
\begin{equation}
\boldsymbol{Y}_{\textrm{S}}=\boldsymbol{A}\left(\boldsymbol{\theta}\right)\boldsymbol{B}\boldsymbol{V}^{T}\left(\boldsymbol{\theta}\right)\boldsymbol{X}+\boldsymbol{Z}_{\textrm{S}},\label{eq:YS}
\end{equation}
where $\boldsymbol{Y}_{\textrm{S}}=\left[\boldsymbol{y}_{\textrm{S}}\left(1\right),\ldots,\boldsymbol{y}_{\textrm{S}}\left(L\right)\right]$,
$\boldsymbol{A}\left(\boldsymbol{\theta}\right)=\left[\boldsymbol{a}\left(\theta_{i}\right),\ldots,\boldsymbol{a}\left(\theta_{K_{\textrm{S}}}\right)\right]$,
$\boldsymbol{V}\left(\boldsymbol{\theta}\right)=\left[\boldsymbol{v}\left(\theta_{i}\right),\ldots,\boldsymbol{v}\left(\theta_{K_{\textrm{S}}}\right)\right]$,
$\boldsymbol{X}=\left[\boldsymbol{x}\left(1\right),\ldots,\boldsymbol{x}\left(L\right)\right]$,
$\boldsymbol{\theta}=\left[\theta_{1},\ldots,\theta_{K_{\textrm{S}}}\right]^{T}$,
$\boldsymbol{b}=\left[\beta_{1},\ldots,\beta_{K_{\textrm{S}}}\right]^{T}$,
and $\boldsymbol{B}=\textrm{diag}\left(\boldsymbol{b}\right)$.

Based on the received echo signal in \eqref{eq:YS}, the BS needs
to estimate the unknow target parameters $\boldsymbol{\theta}$, $\boldsymbol{b}_{\textrm{R}}$,
and $\boldsymbol{b}_{\textrm{I}}$, where $\boldsymbol{b}_{\textrm{R}}=\Re\left(\boldsymbol{b}\right)$
and $\boldsymbol{b}_{\textrm{I}}=\Im\left(\boldsymbol{b}\right)$. 

In the following, to evalute the sensing performance, we adopt the
estimation CRB, which is defined as the lower bound of estimation
error of any unbiased estimators. The Fisher information matrix (FIM)
w.r.t. $\boldsymbol{\theta}$, $\boldsymbol{b}_{\textrm{R}}$, and
$\boldsymbol{b}_{\textrm{I}}$ is given by \cite{4359542}

\begin{align}
\boldsymbol{M} & =\frac{2}{\sigma_{\textrm{S}}^{2}}\begin{bmatrix}\Re\left(\boldsymbol{M}_{11}\right) & \Re\left(\boldsymbol{M}_{12}\right) & -\Im\left(\boldsymbol{M}_{12}\right)\\
\Re^{T}\left(\boldsymbol{M}_{12}\right) & \Re\left(\boldsymbol{M}_{22}\right) & -\Im\left(\boldsymbol{M}_{22}\right)\\
-\Im^{T}\left(\boldsymbol{M}_{12}\right) & -\Im^{T}\left(\boldsymbol{M}_{22}\right) & \Re\left(\boldsymbol{M}_{22}\right)
\end{bmatrix},\label{eq:M_FIM}
\end{align}
where
\begin{align}
\boldsymbol{M}_{11} & =L\left(\dot{\boldsymbol{A}}^{H}\dot{\boldsymbol{A}}\right)\odot\left(\boldsymbol{B}^{c}\boldsymbol{V}^{H}\boldsymbol{R}_{\boldsymbol{x}}^{c}\boldsymbol{V}\boldsymbol{B}\right)\nonumber \\
 & +L\left(\dot{\boldsymbol{A}}^{H}\boldsymbol{A}\right)\odot\left(\boldsymbol{B}^{c}\boldsymbol{V}^{H}\boldsymbol{R}_{\boldsymbol{x}}^{c}\dot{\boldsymbol{V}}\boldsymbol{B}\right)\nonumber \\
 & +L\left(\boldsymbol{A}^{H}\dot{\boldsymbol{A}}\right)\odot\left(\boldsymbol{B}^{c}\dot{\boldsymbol{V}}^{H}\boldsymbol{R}_{\boldsymbol{x}}^{c}\boldsymbol{V}\boldsymbol{B}\right)\nonumber \\
 & +L\left(\boldsymbol{A}^{H}\boldsymbol{A}\right)\odot\left(\boldsymbol{B}^{c}\dot{\boldsymbol{V}}^{H}\boldsymbol{R}_{\boldsymbol{x}}^{c}\dot{\boldsymbol{V}}\boldsymbol{B}\right),\label{eq:M11}
\end{align}
\begin{align}
\boldsymbol{M}_{12} & =L\left(\dot{\boldsymbol{A}}^{H}\boldsymbol{A}\right)\odot\left(\boldsymbol{B}^{c}\boldsymbol{V}^{H}\boldsymbol{R}_{\boldsymbol{x}}^{c}\boldsymbol{V}\right)\nonumber \\
 & +L\left(\boldsymbol{A}^{H}\boldsymbol{A}\right)\odot\left(\boldsymbol{B}^{c}\dot{\boldsymbol{V}}^{H}\boldsymbol{R}_{\boldsymbol{x}}^{c}\boldsymbol{V}\right),\label{eq:M12}
\end{align}
\begin{equation}
\boldsymbol{M}_{22}=L\left(\boldsymbol{A}^{H}\boldsymbol{A}\right)\odot\left(\boldsymbol{V}^{H}\boldsymbol{R}_{\boldsymbol{x}}^{c}\boldsymbol{V}\right),\label{eq:M22}
\end{equation}
\begin{equation}
\dot{\boldsymbol{A}}=\left[\dot{\boldsymbol{a}}\left(\theta_{1}\right),\cdots,\dot{\boldsymbol{a}}\left(\theta_{K_{\textrm{S}}}\right)\right],
\end{equation}
\begin{equation}
\dot{\boldsymbol{V}}=\left[\dot{\boldsymbol{v}}\left(\theta_{1}\right),\cdots,\dot{\boldsymbol{v}}\left(\theta_{K_{\textrm{S}}}\right)\right].
\end{equation}

Consequently, the CRB matrix is given by
\begin{equation}
\textrm{CRB}\left(\boldsymbol{F},\left\{ \boldsymbol{w}_{k}\right\} ,\boldsymbol{S}\right)=\boldsymbol{M}^{-1}.\label{eq:CRB}
\end{equation}
We use the trace of the CRB matrix as the performance metric for sensing,
which is expressed as
\begin{equation}
\overline{\textrm{CRB}}\left(\boldsymbol{F},\left\{ \boldsymbol{w}_{k}\right\} ,\boldsymbol{S}\right)=\textrm{tr}\left(\boldsymbol{M}^{-1}\right).\label{eq:barCRB}
\end{equation}
In \eqref{eq:barCRB}, $\overline{\textrm{CRB}}$ characterizes the
average CRB of all estimated parameters, which is widely adopted as
the sensing performance measure in prior works (e.g., \cite{Zhu2023a}).

\subsection{WPT Model}

In this subsection, we consider the WPT from the BS to the ERs, which
employ rectifiers to convert RF signals to direct current (DC) signals
for energy harvesting. The received RF power at ER $j$ is given by
\begin{align}
P_{j}^{\textrm{EH}} & \left(\boldsymbol{F},\left\{ \boldsymbol{w}_{k}\right\} ,\boldsymbol{S}\right)=\boldsymbol{d}_{j}^{H}\boldsymbol{F}\left(\underset{k\in\mathcal{K}_{\textrm{IR}}}{\sum}\boldsymbol{w}_{k}\boldsymbol{w}_{k}^{H}+\boldsymbol{S}\right)\boldsymbol{F}^{H}\boldsymbol{d}_{j}.
\end{align}
We consider a practical parametric non-linear EH model to characterize
the effect of the non-linear RF-to-DC conversion process, where the
change of RF energy conversion efficiency w.r.t. various input power
levels is captured \cite{7264986}. The total harvested DC power at
ER $j\in\mathcal{K}_{\textrm{ER}}$ is modeled as
\begin{equation}
P_{j}^{\textrm{DC}}\left(\boldsymbol{F},\left\{ \boldsymbol{w}_{k}\right\} ,\boldsymbol{S}\right)=\frac{\varPsi_{j}\left(\boldsymbol{F},\left\{ \boldsymbol{w}_{k}\right\} ,\boldsymbol{S}\right)-M_{j}\varOmega_{j}}{1-\varOmega_{j}},\label{eq:NL_EH_Model}
\end{equation}
with
\begin{equation}
\varOmega_{j}=\frac{1}{1+\textrm{exp}\left(a_{j}b_{j}\right)},
\end{equation}
{\small{}
\begin{align}
 & \varPsi_{j}\left(\boldsymbol{F},\left\{ \boldsymbol{w}_{k}\right\} ,\boldsymbol{S}\right)=\frac{M_{j}}{1+\textrm{exp}\left\{ -a_{j}\left[P_{j}^{\textrm{EH}}\left(\boldsymbol{F},\left\{ \boldsymbol{w}_{k}\right\} ,\boldsymbol{S}\right)-b_{j}\right]\right\} }\nonumber \\
 & =\frac{M_{j}}{1+\textrm{exp}\left\{ -a_{j}\left[\boldsymbol{d}_{j}^{H}\boldsymbol{F}\left(\underset{k\in\mathcal{K}_{\textrm{IR}}}{\sum}\boldsymbol{w}_{k}\boldsymbol{w}_{k}^{H}+\boldsymbol{S}\right)\boldsymbol{F}^{H}\boldsymbol{d}_{j}-b_{j}\right]\right\} },\label{eq:WPT_model}
\end{align}
}where $\varOmega_{j}$ is a constant introduced to ensure zero output
with zero input, $M_{j}$ is a constant representing the saturated
harvested power at ER $j$, while the constants $a_{j}$ and $b_{j}$
are parameters dependent on specific circuits including the capacitance
and resistance.

\subsection{Power Consumption Model}

This subsection presents a comprehensive power consumption model of
the hybrid HAD transmitter and the fully-digital sensing receiver
at the multi-functional BS for facilitating the energy efficient design.
In particular, the power consumption of BS transmitter includes the
non-linear power consumption by PAs, the on-off non-transmission power
consumption by RF chains and PSs, and the static power consumption
by the switch network and other components, while the power consumption
of sensing receiver consists of that consumed by the low noise amplifiers
(LNAs) and other signal processing components \cite{10159012,Skrimponis2020}.

\textit{1) Power Consumption by PAs:} The PA power consumption at
the BS transmitter depends on the output or radiated power at each
transmit antenna as well as the power efficiency. Based on the signal
model in \eqref{eq:x(L)}, the radiated power at each antenna $n$
is given by $P_{n}^{\textrm{out}}=\sum_{k\in\mathcal{K}_{\textrm{IR}}}\left|\boldsymbol{1}_{n}^{N_{\textrm{T}}}\boldsymbol{F}\boldsymbol{w}_{k}\right|^{2}+\textrm{tr}\left(\boldsymbol{E}_{n}^{N_{\textrm{T}}}\boldsymbol{F}\boldsymbol{S}\boldsymbol{F}^{H}\right)$,
where $\boldsymbol{1}_{n}^{N_{\textrm{T}}}\in\mathbb{\mathbb{R}}^{1\times N_{\textrm{T}}}$
is a vector with its $n$-th element being one and others zero, and
$\boldsymbol{E}_{n}^{N_{\textrm{T}}}\in\mathbb{C}^{N_{\textrm{T}}\times N_{\textrm{T}}}$
is a square matrix with the $n$-th diagonal element being 1 and the
others being zero. Furthermore, we consider a practical non-linear
PA efficiency $\eta\left(P_{n}^{\textrm{out}}\right)$, which is modeled
as a non-linear function of the transmit power $P_{n}^{\textrm{out}}$
at antenna $n$, i.e.,
\begin{equation}
\eta\left(P_{n}^{\textrm{out}}\right)=\eta^{\textrm{max}}\left(\frac{P_{n}^{\textrm{out}}}{P_{n}^{\textrm{max}}}\right)^{\beta},0<\beta\leq1,
\end{equation}
where $P_{n}^{\textrm{max}}$ and $P_{n}^{\textrm{out}}$ denote the
maximum and actual transmit power at each antenna $n$, respectively.
In addition, $\eta^{\textrm{max}}$ represents the maximum PA efficiency
under saturation, and $\beta$ signifies the efficiency factor, which
varies based on the specific type of PA \cite{10304516} (e.g., $\beta=1$
and $\beta=0.5$ for class-A and class-B PAs, respectively). Consequently,
the power consumption of all PAs is given by{\small{}
\begin{alignat}{1}
 & P_{\textrm{PA}}\left(\boldsymbol{F},\left\{ \boldsymbol{w}_{k}\right\} ,\boldsymbol{S}\right)=\underset{n\in\mathcal{N}_{\textrm{T}}}{\sum}\frac{1}{\eta}P_{n}^{\textrm{out}}\nonumber \\
 & =\underset{n\in\mathcal{N}_{\textrm{T}}}{\sum}\frac{1}{\eta^{\textrm{max}}}\left(P_{n}^{\textrm{max}}\right)^{\beta}\left(P_{n}^{\textrm{out}}\right)^{1-\beta}\nonumber \\
 & =\frac{\left(P_{n}^{\textrm{max}}\right)^{\beta}}{\eta^{\textrm{max}}}\underset{n\in\mathcal{N}_{\textrm{T}}}{\sum}\left(\underset{k\in\mathcal{K}_{\textrm{IR}}}{\sum}\left|\boldsymbol{1}_{n}^{N_{\textrm{T}}}\boldsymbol{F}\boldsymbol{w}_{k}\right|^{2}+\textrm{tr}\left(\boldsymbol{E}_{n}^{N_{\textrm{T}}}\boldsymbol{F}\boldsymbol{S}\boldsymbol{F}^{H}\right)\right)^{1-\beta}.\label{eq:model_PA}
\end{alignat}
}Furthermore, the transmit power at each antenna should not exceed
the maximum value $P_{n}^{\textrm{max}}$, i.e.,
\begin{align}
 & P_{n}^{\textrm{out}}\left(\boldsymbol{F},\left\{ \boldsymbol{w}_{k}\right\} ,\boldsymbol{S}\right)\nonumber \\
 & =\underset{k\in\mathcal{K}_{\textrm{IR}}}{\sum}\left|\boldsymbol{1}_{n}^{N_{\textrm{T}}}\boldsymbol{F}\boldsymbol{w}_{k}\right|^{2}+\textrm{tr}\left(\boldsymbol{E}_{n}^{N_{\textrm{T}}}\boldsymbol{F}\boldsymbol{S}\boldsymbol{F}^{H}\right)\leq P_{n}^{\textrm{max}}.\label{eq:model_PAoutConstraint}
\end{align}

\textit{2) Power Consumption by RF Chains and PSs with On-off Control:}
Note that the power consumed by RF chains and PSs depends on their
on-off status. Specifically, when each RF chain is turned on, or equivalently,
$\underset{k\in\mathcal{K}_{\textrm{IR}}}{\sum}\left|\boldsymbol{1}_{n}^{N_{\textrm{RF}}}\boldsymbol{w}_{k}\right|^{2}+\textrm{tr}\left(\boldsymbol{E}_{n}^{N_{\textrm{RF}}}\boldsymbol{S}\right)>0$,
it consumes a fixed power

\begin{equation}
P_{\textrm{RF}}^{\textrm{s}}=P_{\textrm{DAC}}+P_{\textrm{mix}}+P_{\textrm{filt}}+P_{\textrm{syn}},
\end{equation}
where $P_{\textrm{DAC}}$, $P_{\textrm{mix}}$, $P_{\textrm{filt}}$,
and $P_{\textrm{syn}}$ represent the power consumption by the digital-to-analog
converter (DAC), mixer, filter, and synthesizer, respectively. Conversely,
when the RF chain is switched off, or equivalently $\underset{k\in\mathcal{K}_{\textrm{IR}}}{\sum}\left|\boldsymbol{1}_{n}^{N_{\textrm{RF}}}\boldsymbol{w}_{k}\right|^{2}+\textrm{tr}\left(\boldsymbol{E}_{n}^{N_{\textrm{RF}}}\boldsymbol{S}\right)=0$,
the power consumption becomes zero. The total power consumption by
all $N_{\textrm{RF}}$ RF chains is given by
\begin{align}
 & P_{\textrm{RF}}\left(\left\{ \boldsymbol{w}_{k}\right\} ,\boldsymbol{S}\right)\nonumber \\
 & =P_{\textrm{RF}}^{\textrm{s}}\underset{n\in\mathcal{N}_{\textrm{RF}}}{\sum}\mathcal{I}\left\{ \underset{k\in\mathcal{K}_{\textrm{IR}}}{\sum}\left|\boldsymbol{1}_{n}^{N_{\textrm{RF}}}\boldsymbol{w}_{k}\right|^{2}+\textrm{tr}\left(\boldsymbol{E}_{n}^{N_{\textrm{RF}}}\boldsymbol{S}\right)\right\} .\label{eq:model_RF}
\end{align}

Similar to RF chains, an active PS consumes a constant power value
of $P_{\textrm{PS}}^{\textrm{s}}$ when it is activated and zero power
when it is turned off. The total power consumption by PSs is expressed
as 
\begin{equation}
P_{\textrm{PS}}\left(\boldsymbol{F}\right)=P_{\textrm{PS}}^{\textrm{s}}\underset{i\in\mathcal{N}_{\textrm{T}}}{\sum}\underset{j\in\mathcal{N}_{\textrm{RF}}}{\sum}\mathcal{I}\left\{ \left[\boldsymbol{F}\right]_{i,j}\right\} .\label{eq:model_PS}
\end{equation}

\textit{3) Power Consumption by the Switch Network: }Notice that the
switch network also consumes certain energy. Let $P_{\textrm{SW}}^{\textrm{s}}$
denote the power consumption by a single switch, which is typically
much lower than that by a PS. As there are a total of $N_{\textrm{RF}}+N_{\textrm{T}}N_{\textrm{RF}}$
switches that are consistently active, the total power consumption
by the switch network is given by a constant term
\begin{equation}
P_{\textrm{SW}}=P_{\textrm{SW}}^{\textrm{s}}\left(N_{\textrm{RF}}+N_{\textrm{T}}N_{\textrm{RF}}\right).\label{eq:model_Switch}
\end{equation}

\textit{4) Static Power Consumption: }Besides the above parts, other
components such as the power supply, backhaul, and baseband processing
at the BS also introduce certain energy consumption, which is typically
constant, denoted as $P_{\textrm{static}}$.

By combining the power consumptions proposed above, we have the total
power consumption by the BS as

\begin{align}
P_{\textrm{BS}}^{\textrm{tot}}\left(\boldsymbol{F},\left\{ \boldsymbol{w}_{k}\right\} ,\boldsymbol{S}\right) & =P_{\textrm{PA}}\left(\boldsymbol{F},\left\{ \boldsymbol{w}_{k}\right\} ,\boldsymbol{S}\right)+P_{\textrm{RF}}\left(\left\{ \boldsymbol{w}_{k}\right\} ,\boldsymbol{S}\right)\nonumber \\
 & +P_{\textrm{PS}}\left(\boldsymbol{F}\right)+P_{\textrm{SW}}+P_{\textrm{static}}.
\end{align}

\subsection{Problem Formulation}

Our objective is to optimize the joint hybrid beamforming and on-off
control to minimize the total power consumption $P_{\textrm{BS}}^{\textrm{tot}}\left(\boldsymbol{F},\left\{ \boldsymbol{w}_{k}\right\} ,\boldsymbol{S}\right)$
while adhering to specific constraints, including the minimum SINR
constraints $\Gamma_{k}^{\textrm{IR}},\forall k\in\mathcal{K}_{\textrm{IR}},$
at IRs for communication in \eqref{eq:P1_Communication}, the maximum
estimation CRB constraint $\Gamma_{\textrm{S}}$ for target sensing
in \eqref{eq:P1_Sensing}, the minimum harvested DC power constraints
$\Gamma_{j}^{\textrm{DC}},\forall j\in\mathcal{K}_{\textrm{ER}},$
at ERs for WPT in \eqref{eq:P1_Powering}, the per-antenna transmit
power constraint $P_{n}^{\textrm{max}}$ in \eqref{eq:P1_PerAntennaPower},
and the constant modulus constraint for analog beamforming in \eqref{eq:P1_ConstantModulus}.
Notice that the EH constraints are reduced to the form in \eqref{eq:P1_Powering}
by letting $\tilde{\Gamma}_{j}^{\textrm{EH}}=b_{j}-\frac{1}{a_{j}}\ln\left[\frac{M_{j}}{\Gamma_{j}^{\textrm{DC}}\left(1-\varOmega_{j}\right)+M_{j}\varOmega_{j}}-1\right]$,
where $\Gamma_{j}^{\textrm{DC}},\forall j\in\mathcal{K}_{\textrm{ER}},$
denotes the constraint on the harvested DC power $P_{j}^{\textrm{DC}}\left(\boldsymbol{F},\left\{ \boldsymbol{w}_{k}\right\} ,\boldsymbol{S}\right)$
at ER $j$, according to \eqref{eq:WPT_model}. As a result, the problem
is formulated as (P1), shown at the top of the next page. The BS is
assumed to have perfect knowledge of channel state information (CSI)
as well as rough prior information of the estimated parameters to
facilitate wireless communication and target tracking.
\begin{figure*}[tbh]
\begin{eqnarray}
\left(\mathrm{P1}\right): & \underset{\boldsymbol{F},\left\{ \boldsymbol{w}_{k}\right\} ,\boldsymbol{S}\succeq0}{\min} & \underset{P_{\textrm{PA}}\left(\boldsymbol{F},\left\{ \boldsymbol{w}_{k}\right\} ,\boldsymbol{S}\right)}{\underbrace{\frac{\left(P_{n}^{\textrm{max}}\right)^{\beta}}{\eta^{\textrm{max}}}\underset{n\in\mathcal{N}_{\textrm{T}}}{\sum}\left(\underset{k\in\mathcal{K}_{\textrm{IR}}}{\sum}\left|\boldsymbol{1}_{n}^{N_{\textrm{T}}}\boldsymbol{F}\boldsymbol{w}_{k}\right|^{2}+\textrm{tr}\left(\boldsymbol{E}_{n}^{N_{\textrm{T}}}\boldsymbol{F}\boldsymbol{S}\boldsymbol{F}^{H}\right)\right)^{1-\beta}}}\nonumber \\
 &  & +\underset{P_{\textrm{RF}}\left(\left\{ \boldsymbol{w}_{k}\right\} ,\boldsymbol{S}\right)}{\underbrace{P_{\textrm{RF}}^{\textrm{s}}\underset{n\in\mathcal{N}_{\textrm{RF}}}{\sum}\mathcal{I}\left\{ \underset{k\in\mathcal{K}_{\textrm{IR}}}{\sum}\left|\boldsymbol{1}_{n}^{N_{\textrm{RF}}}\boldsymbol{w}_{k}\right|^{2}+\textrm{tr}\left(\boldsymbol{E}_{n}^{N_{\textrm{RF}}}\boldsymbol{S}\right)\right\} }}+\underset{P_{\textrm{PS}}\left(\boldsymbol{F}\right)}{\underbrace{P_{\textrm{PS}}^{\textrm{s}}\underset{i\in\mathcal{N}_{\textrm{T}}}{\sum}\underset{j\in\mathcal{N}_{\textrm{RF}}}{\sum}\mathcal{I}\left\{ \left[\boldsymbol{F}\right]_{i,j}\right\} }}+P_{\textrm{SW}}+P_{\textrm{static}}\nonumber \\
 & \mathrm{s.t.} & \frac{\left|\boldsymbol{h}_{k}^{H}\boldsymbol{F}\boldsymbol{w}_{k}\right|^{2}}{\underset{i\in\mathcal{K}_{\textrm{IR}},i\neq k}{\sum}\left|\boldsymbol{h}_{k}^{H}\boldsymbol{F}\boldsymbol{w}_{i}\right|^{2}+\left|\boldsymbol{h}_{k}^{H}\boldsymbol{F}\boldsymbol{S}\boldsymbol{F}^{H}\boldsymbol{h}_{k}\right|+\sigma_{k}^{2}}\geq\Gamma_{k}^{\textrm{IR}},\forall k\in\mathcal{K}_{\textrm{IR}},\label{eq:P1_Communication}\\
 &  & \overline{\textrm{CRB}}\left(\boldsymbol{F},\left\{ \boldsymbol{w}_{k}\right\} ,\boldsymbol{S}\right)\leq\Gamma_{\textrm{S}},\label{eq:P1_Sensing}\\
 &  & \boldsymbol{d}_{j}^{H}\boldsymbol{F}\left(\underset{k\in\mathcal{K}_{\textrm{IR}}}{\sum}\boldsymbol{w}_{k}\boldsymbol{w}_{k}^{H}+\boldsymbol{S}\right)\boldsymbol{F}^{H}\boldsymbol{d}_{j}\geq\tilde{\Gamma}_{j}^{\textrm{EH}},\forall j\in\mathcal{K}_{\textrm{ER}},\label{eq:P1_Powering}\\
 &  & \underset{k\in\mathcal{K}_{\textrm{IR}}}{\sum}\left|\boldsymbol{1}_{n}^{N_{\textrm{T}}}\boldsymbol{F}\boldsymbol{w}_{k}\right|^{2}+\textrm{tr}\left(\boldsymbol{E}_{n}^{N_{\textrm{T}}}\boldsymbol{F}\boldsymbol{S}\boldsymbol{F}^{H}\right)\leq P_{n}^{\textrm{max}},\forall n\in\mathcal{N}_{\textrm{T}},\label{eq:P1_PerAntennaPower}\\
 &  & \left[\boldsymbol{F}\right]_{i,j}\in\mathcal{F},\forall i\in\mathcal{N}_{\textrm{T}},j\in\mathcal{N}_{\textrm{RF}}.\label{eq:P1_ConstantModulus}
\end{eqnarray}
\hrule
\end{figure*}

The formulated problem exhibits a highly non-convex nature, primarily
stemming from the coupling between analog and digital beamformers,
the elementwise constant modulus constraint for the analog beamformer
in \eqref{eq:P1_ConstantModulus}, the quadratic objective and constraints,
and the non-linear PA efficiency and the binary on-off power consumption
of RF chains and PSs in the power consumption model. Due to such difficulties,
this problem is extremely challenging and thus has not been studied
in the existing literature.

\section{Proposed Solution to Problem (P1) Based on Alternating Optimization}

This section proposes an efficient algorithm to address problem (P1)
by employing the AO technique. In particular, we alternatively optimize
the digital beamforming $\left\{ \boldsymbol{w}_{k}\right\} $ and
$\boldsymbol{S}$, and the analog beamforming $\boldsymbol{F}$. Based
on the obtained beamforming solution, we propose a dynamic on-off
switching algorithm to determine the on-off statuses of RF chains
and PSs for reducing the total power consumption.

To begin with, we approximate the binary on-off non-transmission power
in \eqref{eq:model_RF} and \eqref{eq:model_PS} into continuous forms,
by considering the following approximation for the indicator function
\cite{Sriperumbudur2011}

\begin{equation}
\mathcal{I}\left\{ x\right\} =\underset{\varepsilon\rightarrow0}{\lim}\frac{\log\left(1+x\varepsilon^{-1}\right)}{\log\left(1+\varepsilon^{-1}\right)},x\geq0.\label{eq:Ind_APPRO}
\end{equation}
Accordingly, the indicator function is approximated as $\frac{\log\left(1+x\varepsilon^{-1}\right)}{\log\left(1+\varepsilon^{-1}\right)}$
with a properly chosen parameter $\varepsilon$, which is small enough
to guarantee the accuracy of the approximation. By using \eqref{eq:Ind_APPRO},
the total power consumption by RF chains and PSs are respectively
approximated as

\begin{align}
 & \hat{P}_{\mathrm{RF}}\left(\left\{ \boldsymbol{w}_{k}\right\} ,\boldsymbol{S}\right)=\frac{P_{\textrm{RF}}^{\textrm{s}}}{\log\left(1+\varepsilon^{-1}\right)}\times\nonumber \\
 & \underset{n\in\mathcal{N}_{\textrm{RF}}}{\sum}\log\left\{ 1+\varepsilon^{-1}\left[\underset{k\in\mathcal{K}_{\textrm{IR}}}{\sum}\left|\boldsymbol{1}_{n}^{N_{\textrm{RF}}}\boldsymbol{w}_{k}\right|^{2}+\textrm{tr}\left(\boldsymbol{E}_{n}^{N_{\textrm{RF}}}\boldsymbol{S}\right)\right]\right\} ,\label{eq:RF_I_appro-1}
\end{align}
\begin{align}
\hat{P}_{\textrm{PS}}\left(\boldsymbol{F}\right) & =\frac{P_{\textrm{PS}}^{\textrm{s}}}{\log\left(1+\varepsilon^{-1}\right)}\times\nonumber \\
 & \underset{i\in\mathcal{N}_{\textrm{T}}}{\sum}\underset{j\in\mathcal{N}_{\textrm{RF}}}{\sum}\log\left\{ 1+\varepsilon^{-1}\left[\boldsymbol{F}\right]_{i,j}\right\} .\label{eq:PS_I_appro-1}
\end{align}
Thus, by omitting the constant terms $P_{\textrm{SW}}$ and $P_{\textrm{static}}$,
we approximate problem (P1) as
\begin{eqnarray*}
\left(\mathrm{P2}\right): & \underset{\boldsymbol{F},\left\{ \boldsymbol{w}_{k}\right\} ,\boldsymbol{S}\succeq0}{\min} & P_{\textrm{PA}}\left(\boldsymbol{F},\left\{ \boldsymbol{w}_{k}\right\} ,\boldsymbol{S}\right)+\hat{P}_{\textrm{RF}}\left(\left\{ \boldsymbol{w}_{k}\right\} ,\boldsymbol{S}\right)\\
 &  & +\hat{P}_{\textrm{PS}}\left(\boldsymbol{F}\right)\\
 & \mathrm{s.t.} & \eqref{eq:P1_Communication}-\eqref{eq:P1_ConstantModulus}.
\end{eqnarray*}

\subsection{Optimization of Digital Beamformer and Sensing Covariance in (P2)
with Given Analog Beamformer}

We first optimize the digital beamformer $\left\{ \boldsymbol{w}_{k}\right\} $
and sensing covariance matrix $\boldsymbol{S}$ in (P2) under given
analog beamformer $\boldsymbol{F}$. In this case, problem (P2) is
reduced into

\begin{eqnarray*}
\left(\mathrm{P3}\right): & \underset{\left\{ \boldsymbol{w}_{k}\right\} ,\boldsymbol{S}\succeq0}{\min} & P_{\textrm{PA}}\left(\left\{ \boldsymbol{w}_{k}\right\} ,\boldsymbol{S}\right)+\hat{P}_{\textrm{RF}}\left(\left\{ \boldsymbol{w}_{k}\right\} ,\boldsymbol{S}\right)+\hat{P}_{\textrm{PS}}\\
 & \mathrm{s.t.} & \eqref{eq:P1_Communication}-\eqref{eq:P1_PerAntennaPower}.
\end{eqnarray*}

Note that problem (P3) is still non-convex due to the quadratic terms
of $\left\{ \boldsymbol{w}_{k}\right\} $ in the objective function
and the constraints as well as the non-linear objective function.
We first use the SDR technique to deal with the quadratic forms of
$\left\{ \boldsymbol{w}_{k}\right\} $. Specifically, by letting $\boldsymbol{R}_{k}=\boldsymbol{w}_{k}\boldsymbol{w}_{k}^{H}$
with $\boldsymbol{R}_{k}\succeq0$ and $\mathrm{rank}\left(\boldsymbol{R}_{k}\right)=1$,
the PA and RF chain power consumption terms in the objective function
are rewritten as
\begin{align}
 & \bar{P}_{\textrm{PA}}\left(\left\{ \boldsymbol{R}_{k}\right\} ,\boldsymbol{S}\right)=\frac{1}{\eta^{\textrm{max}}}\underset{n\in\mathcal{N}_{\textrm{T}}}{\sum}\left(P_{n}^{\textrm{max}}\right)^{\beta}\times\nonumber \\
 & \left\{ \textrm{tr}\left[\boldsymbol{E}_{n}^{N_{\textrm{T}}}\boldsymbol{F}\left(\underset{k\in\mathcal{K}_{\textrm{IR}}}{\sum}\boldsymbol{R}_{k}+\boldsymbol{S}\right)\boldsymbol{F}^{H}\right]\right\} ^{1-\beta},\label{eq:PA_Rk_S}
\end{align}
\begin{align}
 & \bar{P}_{\textrm{RF}}\left(\left\{ \boldsymbol{R}_{k}\right\} ,\boldsymbol{S}\right)=\frac{P_{\textrm{RF}}^{\textrm{s}}}{\log\left(1+\varepsilon^{-1}\right)}\times\nonumber \\
 & \underset{n\in\mathcal{N}_{\textrm{RF}}}{\sum}\log\left\{ 1+\varepsilon^{-1}\textrm{tr}\left[\boldsymbol{E}_{n}^{N_{\textrm{RF}}}\left(\underset{k\in\mathcal{K}_{\textrm{IR}}}{\sum}\boldsymbol{R}_{k}+\boldsymbol{S}\right)\right]\right\} .\label{eq:RF_I_appro-2}
\end{align}
 respectively. Notice that \eqref{eq:P1_ConstantModulus} and \eqref{eq:RF_I_appro-2}
are both concave functions of $\left\{ \boldsymbol{R}_{k}\right\} $
and $\boldsymbol{S}$. Upon letting $\boldsymbol{H}_{k}=\boldsymbol{h}_{k}\boldsymbol{h}_{k}^{H}$,
the constraints in \eqref{eq:P1_Communication} are equivalent to
\begin{align}
\textrm{tr}\left(\boldsymbol{H}_{k}\boldsymbol{F}\boldsymbol{R}_{k}\boldsymbol{F}^{H}\right)-\Gamma_{\textrm{IR}}\underset{i\in\mathcal{K}_{\textrm{IR}},i\neq k}{\sum}\textrm{tr}\left(\boldsymbol{H}_{k}\boldsymbol{F}\boldsymbol{R}_{i}\boldsymbol{F}^{H}\right)\nonumber \\
-\Gamma_{\textrm{IR}}\textrm{tr}\left(\boldsymbol{H}_{k}\boldsymbol{F}\boldsymbol{S}\boldsymbol{F}^{H}\right)\geq\Gamma_{k}^{\textrm{IR}}\sigma_{k}^{2},\forall k\in\mathcal{K}_{\textrm{IR}},\label{eq:P1_Communication_SDR}
\end{align}
which are convex w.r.t. $\left\{ \boldsymbol{R}_{k}\right\} $ and
$\boldsymbol{S}$. Then, by applying the Schur complement, the constraint
in \eqref{eq:P1_Sensing} is reformulated into the following convex
form
\begin{equation}
\begin{cases}
\stackrel[i=1]{3K_{\textrm{S}}}{\sum}t_{i}\leq\Gamma_{\textrm{S}},\\
\begin{bmatrix}\boldsymbol{M} & \boldsymbol{e}_{i}\\
\boldsymbol{e}_{i}^{T} & t_{i}
\end{bmatrix}\succeq0,i=1,\ldots,3K_{\textrm{S}},
\end{cases}\label{eq:P1_Sensing_PreProcess}
\end{equation}
where $\boldsymbol{e}_{i}\in\mathbb{R}^{3K_{\textrm{S}}\times1}$
is the $i$-th column of the identity matrix $\boldsymbol{I}_{3K_{\textrm{S}}}$,
and $\left\{ t_{i}\right\} $ are newly introduced slack variables.
The constraints in \eqref{eq:P1_Powering} are equivalent to 
\begin{equation}
\boldsymbol{d}_{j}^{H}\boldsymbol{F}\left(\underset{k\in\mathcal{K}_{\textrm{IR}}}{\sum}\boldsymbol{R}_{k}+\boldsymbol{S}\right)\boldsymbol{F}^{H}\boldsymbol{d}_{j}\geq\tilde{\Gamma}_{j}^{\textrm{EH}},\forall j\in\mathcal{K}_{\textrm{ER}},\label{eq:P1_Powering_PreProcess}
\end{equation}
which are convex w.r.t. $\left\{ \boldsymbol{R}_{k}\right\} $ and
$\boldsymbol{S}$. Furthermore, the constraints in \eqref{eq:P1_PerAntennaPower}
are reformulated into
\begin{equation}
\textrm{tr}\left[\boldsymbol{E}_{n}^{N_{\textrm{T}}}\boldsymbol{F}\left(\underset{k\in\mathcal{K}_{\textrm{IR}}}{\sum}\boldsymbol{R}_{k}+\boldsymbol{S}\right)\boldsymbol{F}^{H}\right]\leq P_{n}^{\textrm{max}},\forall n\in\mathcal{N}_{\textrm{T}},\label{eq:P1_Per_Antenna_PreProcess}
\end{equation}
which are also convex w.r.t. $\left\{ \boldsymbol{R}_{k}\right\} $
and $\boldsymbol{S}$. Thanks to the above derivations, both the objective
function and the constraints are recast as functions of $\left\{ \boldsymbol{R}_{k}\right\} $
and $\boldsymbol{S}$. As a result, problem (P3) is approximated into
problem (SDR3) by relaxing the rank-one constraint $\mathrm{rank}\left(\boldsymbol{R}_{k}\right)=1$,
where $\left\{ \boldsymbol{R}_{k}\right\} $ and $\boldsymbol{S}$
are optimized instead of $\left\{ \boldsymbol{w}_{k}\right\} $ and
$\boldsymbol{S}$.
\begin{eqnarray*}
\left(\mathrm{SDR3}\right): & \underset{\left\{ t_{i}\right\} ,\left\{ \boldsymbol{R}_{k}\right\} \succeq0,\boldsymbol{S}\succeq0}{\min} & \bar{P}_{\textrm{PA}}\left(\left\{ \boldsymbol{R}_{k}\right\} ,\boldsymbol{S}\right)\\
 &  & +\bar{P}_{\textrm{RF}}\left(\left\{ \boldsymbol{R}_{k}\right\} ,\boldsymbol{S}\right)+\hat{P}_{\textrm{PS}}\\
 & \mathrm{s.t.} & \eqref{eq:P1_Communication_SDR}-\eqref{eq:P1_Per_Antenna_PreProcess}.
\end{eqnarray*}

Then, we utilize the SCA technique to address the non-convex objective
function in problem (SDR3). Let $\left\{ \boldsymbol{R}_{k}^{\left(j\right)}\right\} $
and $\boldsymbol{S}^{\left(j\right)}$ represent the local points
of $\left\{ \boldsymbol{R}_{k}\right\} $ and $\boldsymbol{S}$ in
a specific SCA iteration $j$, respectively. By using the first-order
Taylor expansion, we approximate the PA and RF chain power consumption
terms $\bar{P}_{\textrm{PA}}\left(\left\{ \boldsymbol{R}_{k}\right\} ,\boldsymbol{S}\right)$
and $\bar{P}_{\textrm{RF}}\left(\left\{ \boldsymbol{R}_{k}\right\} ,\boldsymbol{S}\right)$
as their upper bounds $\tilde{P}_{\textrm{PA}}^{\left(j\right)}\left(\left\{ \boldsymbol{R}_{k}\right\} ,\boldsymbol{S}\right)$
in \eqref{eq:SCA_PA_W} and $\tilde{P}_{\textrm{RF}}^{\left(j\right)}\left(\left\{ \boldsymbol{R}_{k}\right\} ,\boldsymbol{S}\right)$
in \eqref{eq:SCA_RF}, respectively, as shown at the top of the next
page. 

\begin{figure*}[tbh]
\begin{align}
\tilde{P}_{\textrm{PA}}^{\left(j\right)}\left(\left\{ \boldsymbol{R}_{k}\right\} ,\boldsymbol{S}\right) & =\underset{n\in\mathcal{N}_{\textrm{T}}}{\sum}\frac{\left(P_{n}^{\textrm{max}}\right)^{\beta}}{\eta^{\textrm{max}}}\left\{ \textrm{tr}\left[\boldsymbol{E}_{n}^{N_{\textrm{T}}}\boldsymbol{F}\left(\underset{k\in\mathcal{K}_{\textrm{IR}}}{\sum}\boldsymbol{R}_{k}^{\left(j\right)}+\boldsymbol{S}^{\left(j\right)}\right)\boldsymbol{F}^{H}\right]\right\} ^{1-\beta}+\left(1-\beta\right)\underset{n\in\mathcal{N}_{\textrm{T}}}{\sum}\frac{\left(P_{n}^{\textrm{max}}\right)^{\beta}}{\eta^{\textrm{max}}}\times\nonumber \\
 & \left\{ \textrm{tr}\left[\boldsymbol{E}_{n}^{N_{\textrm{T}}}\boldsymbol{F}\left(\underset{k\in\mathcal{K}_{\textrm{IR}}}{\sum}\boldsymbol{R}_{k}^{\left(j\right)}+\boldsymbol{S}^{\left(j\right)}\right)\boldsymbol{F}^{H}\right]\right\} ^{-\beta}\textrm{tr}\left\{ \boldsymbol{E}_{n}^{N_{\textrm{T}}}\boldsymbol{F}\left[\underset{k\in\mathcal{K}_{\textrm{IR}}}{\sum}\left(\boldsymbol{R}_{k}-\boldsymbol{R}_{k}^{\left(j\right)}\right)+\left(\boldsymbol{S}-\boldsymbol{S}^{\left(j\right)}\right)\right]\boldsymbol{F}^{H}\right\} \label{eq:SCA_PA_W}
\end{align}
\begin{align}
\tilde{P}_{\textrm{RF}}^{\left(j\right)}\left(\left\{ \boldsymbol{R}_{k}\right\} ,\boldsymbol{S}\right) & =\frac{P_{\textrm{RF}}^{\textrm{s}}}{\log\left(1+\varepsilon^{-1}\right)}\underset{n\in\mathcal{N}_{\textrm{RF}}}{\sum}\left\{ \log\left\{ 1+\varepsilon^{-1}\textrm{tr}\left[\boldsymbol{E}_{n}^{N_{\textrm{RF}}}\left(\underset{k\in\mathcal{K}_{\textrm{IR}}}{\sum}\boldsymbol{R}_{k}^{\left(j\right)}+\boldsymbol{S}^{\left(j\right)}\right)\right]\right\} \right.\nonumber \\
 & \left.+\frac{1}{\textrm{tr}\left[\boldsymbol{E}_{n}^{N_{\textrm{RF}}}\left(\underset{k\in\mathcal{K}_{\textrm{IR}}}{\sum}\boldsymbol{R}_{k}^{\left(j\right)}+\boldsymbol{S}^{\left(j\right)}\right)\right]+\varepsilon}\left\{ \textrm{tr}\left[\boldsymbol{E}_{n}^{N_{\textrm{RF}}}\left(\underset{k\in\mathcal{K}_{\textrm{IR}}}{\sum}\left(\boldsymbol{R}_{k}-\boldsymbol{R}_{k}^{\left(j\right)}\right)+\left(\boldsymbol{S}-\boldsymbol{S}^{\left(j\right)}\right)\right)\right]\right\} \right\} \label{eq:SCA_RF}
\end{align}

\hrule
\end{figure*}

Accordingly, the optimization problem in each iteration $j$ in SCA
is given by
\begin{eqnarray*}
\left(\mathrm{SCA3}.\ensuremath{j}\right): & \underset{\left\{ t_{i}\right\} ,\left\{ \boldsymbol{R}_{k}\right\} \succeq0,\boldsymbol{S}\succeq0}{\min} & \tilde{P}_{\textrm{PA}}^{\left(j\right)}\left(\left\{ \boldsymbol{R}_{k}\right\} ,\boldsymbol{S}\right)\\
 &  & +\tilde{P}_{\textrm{RF}}^{\left(j\right)}\left(\left\{ \boldsymbol{R}_{k}\right\} ,\boldsymbol{S}\right)+\hat{P}_{\textrm{PS}}\\
 & \mathrm{s.t.} & \eqref{eq:P1_Communication_SDR}-\eqref{eq:P1_Per_Antenna_PreProcess}.
\end{eqnarray*}
Note that problem (SCA3.\textit{j}) is a convex problem that can be
efficiently solved by using standard convex optimization tools like
CVX \cite{CVX}. Let $\left\{ \boldsymbol{R}_{k}^{*\left(j\right)}\right\} $
and $\boldsymbol{S}^{*\left(j\right)}$ denote the optimal solutions
to problem (SCA3.\textit{j}), which are used as the local points in
the next iteration. Note that $\tilde{P}_{\textrm{PA}}^{\left(j\right)}\left(\left\{ \boldsymbol{R}_{k}\right\} ,\boldsymbol{S}\right)$
in \eqref{eq:SCA_PA_W}, and $\tilde{P}_{\textrm{RF}}^{\left(j\right)}\left(\left\{ \boldsymbol{R}_{k}\right\} ,\boldsymbol{S}\right)$
in \eqref{eq:SCA_RF} serve as the upper bounds for $\bar{P}_{\textrm{PA}}\left(\left\{ \boldsymbol{R}_{k}\right\} ,\boldsymbol{S}\right)$
in \eqref{eq:PA_Rk_S} and $\bar{P}_{\textrm{RF}}\left(\left\{ \boldsymbol{R}_{k}\right\} ,\boldsymbol{S}\right)$
in \eqref{eq:RF_I_appro-2}, respectively, which imply that the optimal
objective value for problem (SCA3.\textit{j}) in the $\left(j+1\right)$-th
iteration is consistently no greater than that in the $j$-th iteration.
As a result, the achieved power consumption value of problem (SDR3)
diminishes monotonically with the iteration of SCA, thus ensuring
the convergence of the proposed algorithm. Let $\left\{ \boldsymbol{\bar{R}}_{k}^{*}\right\} $
and $\boldsymbol{\bar{S}}^{*}$ denote the converged solution for
communication and sensing covariance, where $\left\{ \boldsymbol{\bar{R}}_{k}^{*}\right\} $
is typically of high rank.

Based on $\left\{ \boldsymbol{\bar{R}}_{k}^{*}\right\} $ and $\boldsymbol{\bar{S}}^{*}$,
we can reconstruct the optimal rank-one solutions of $\left\{ \boldsymbol{R}_{k}\right\} $
to problem (SDR3) and, consequently, for (P3), by using the following
proposition.
\begin{prop}
\textup{The optimal solution to problems (SDR3) and (P3) is given
by}
\end{prop}
\begin{align}
\boldsymbol{R}_{k}^{*} & =\boldsymbol{w}_{k}^{*}\left(\boldsymbol{w}_{k}^{*}\right)^{H},\label{eq:Rank-1-Wk}\\
\boldsymbol{S}^{*} & =\boldsymbol{\bar{S}}^{*}+\underset{k\in\mathcal{K}_{\textrm{IR}}}{\sum}\boldsymbol{\bar{R}}_{k}^{*}-\underset{k\in\mathcal{K}_{\textrm{IR}}}{\sum}\boldsymbol{w}_{k}^{*}\left(\boldsymbol{w}_{k}^{*}\right)^{H},\label{eq:Rank-1-S}
\end{align}
where 
\begin{equation}
\boldsymbol{w}_{k}^{*}=\left(\boldsymbol{h}_{k}^{H}\boldsymbol{F}\boldsymbol{\bar{R}}_{k}^{*}\boldsymbol{F}^{H}\boldsymbol{h}_{k}\right)^{-\frac{1}{2}}\boldsymbol{\bar{R}}_{k}^{*}\boldsymbol{F}^{H}\boldsymbol{h}_{k},\label{eq:Rank-1-wk}
\end{equation}
denotes the corresponding digital beamforming vector.
\begin{IEEEproof}
See Appendix A.
\end{IEEEproof}

\subsection{Optimization of Analog Beamformer in (P2) with Given Digital Beamformer
and Sensing Covariance}

In this subsection, we optimize the analog beamformer $\boldsymbol{F}$
in (P2) under given $\left\{ \boldsymbol{w}_{k}\right\} $ and $\boldsymbol{S}$.
In this case, problem (P2) is reduced into

\begin{eqnarray*}
\left(\mathrm{P4}\right): & \underset{\boldsymbol{F}}{\min} & P_{\textrm{PA}}\left(\boldsymbol{F}\right)+\hat{P}_{\textrm{RF}}+\hat{P}_{\textrm{PS}}\left(\boldsymbol{F}\right)\\
 & \mathrm{s.t.} & \eqref{eq:P1_Communication}-\eqref{eq:P1_ConstantModulus}
\end{eqnarray*}

Problem (P4) is rather challenging due to the quadratic terms of $\boldsymbol{F}$
in the objective function and constraints \eqref{eq:P1_Communication}-\eqref{eq:P1_PerAntennaPower},
the non-linearity of PA power consumption term in the objective function,
as well as the elementwise constant modulus nature of $\boldsymbol{F}$.
To address this problem, we first apply the SDR technique to transform
the quadratic forms of $\boldsymbol{F}$. Let $\boldsymbol{f}=\mathrm{vec}\left(\boldsymbol{F}^{T}\right)$
denote the column vectorization of $\boldsymbol{F}^{T}$, and define
$\boldsymbol{R}_{\boldsymbol{f}}=\boldsymbol{f}\boldsymbol{f}^{H}$
with $\boldsymbol{R}_{\boldsymbol{f}}\succeq0$ and $\mathrm{rank}\left(\boldsymbol{R}_{\boldsymbol{f}}\right)=1$.
As such, we have $\boldsymbol{F}\boldsymbol{w}_{k}=\boldsymbol{E}\left[\boldsymbol{I}^{N_{\textrm{T}}}\otimes\textrm{diag}\left(\boldsymbol{w}_{k}\right)\boldsymbol{f}\right]$,
where $\boldsymbol{E}\in\mathbb{C}^{N_{\textrm{T}}\times N_{\textrm{T}}N_{\textrm{RF}}}$
is defined as 
\begin{equation}
\boldsymbol{E}=\begin{bmatrix}\boldsymbol{e}_{0} & \boldsymbol{0} & \cdots & \boldsymbol{0}\\
\boldsymbol{0} & \boldsymbol{e}_{0} &  & \vdots\\
\vdots &  & \ddots & \boldsymbol{0}\\
\boldsymbol{0} & \cdots & \boldsymbol{0} & \boldsymbol{e}_{0}
\end{bmatrix},
\end{equation}
where $\boldsymbol{e}_{0}\in\mathbb{C}^{1\times N_{\textrm{RF}}}$
is a vector with all its elements being one. Thus, we define
\begin{align}
 & \bar{\boldsymbol{R}}_{k}\left(\boldsymbol{R}_{\boldsymbol{f}}\right)=\left(\boldsymbol{F}\boldsymbol{w}_{k}\right)\left(\boldsymbol{F}\boldsymbol{w}_{k}\right)^{H}\nonumber \\
 & =\boldsymbol{E}\left[\boldsymbol{I}^{N_{\textrm{T}}}\otimes\textrm{diag}\left(\boldsymbol{w}_{k}\right)\right]\boldsymbol{R}_{\boldsymbol{f}}\left[\boldsymbol{I}^{N_{\textrm{T}}}\otimes\textrm{diag}\left(\boldsymbol{w}_{k}^{c}\right)\right]\boldsymbol{E}^{T},\label{eq:Rk(Rf)}
\end{align}
which is a linear function of $\boldsymbol{R}_{\boldsymbol{f}}$.
Let $r=\mathrm{rank}\left(\boldsymbol{S}\right)$, and $\boldsymbol{S}=\sum_{i=1}^{r}\lambda_{i}\boldsymbol{q}_{i}\boldsymbol{q}_{i}^{H}$
denote the eigen-decomposition of the positive semi-definite matrix
$\boldsymbol{S}$, where $\lambda_{1}\geq\lambda_{2}\geq\ldots\geq\lambda_{r}\geq0$
are the eigenvalues and $\boldsymbol{q}_{1},\ldots,\boldsymbol{q}_{r}$
are the respective eigenvectors. We define
\begin{align}
\boldsymbol{R}_{\textrm{S}}\left(\boldsymbol{R}_{\boldsymbol{f}}\right) & =\boldsymbol{F}\boldsymbol{S}\boldsymbol{F}^{H}=\stackrel[i=1]{r}{\sum}\lambda_{i}\left(\boldsymbol{F}\boldsymbol{q}_{i}\right)\left(\boldsymbol{F}\boldsymbol{q}_{i}\right)^{H}\nonumber \\
 & =\stackrel[i=1]{r}{\sum}\lambda_{i}\boldsymbol{E}\left[\boldsymbol{I}^{N_{\textrm{T}}}\otimes\textrm{diag}\left(\boldsymbol{q}_{i}\right)\right]\boldsymbol{R}_{\boldsymbol{f}}\nonumber \\
 & \times\left[\boldsymbol{I}^{N_{\textrm{T}}}\otimes\textrm{diag}\left(\boldsymbol{q}_{i}^{c}\right)\right]\boldsymbol{E}^{T},\label{eq:Rs(Rf)}
\end{align}
which is a linear function of $\boldsymbol{R}_{\boldsymbol{f}}$.
With the help of \eqref{eq:Rk(Rf)} and \eqref{eq:Rs(Rf)}, the PA
and PS power consumption terms in the objective function are re-expressed
as
\begin{align}
\bar{P}_{\textrm{PA}}\left(\boldsymbol{R}_{\boldsymbol{f}}\right) & =\frac{1}{\eta^{\textrm{max}}}\underset{n\in\mathcal{N}_{\textrm{T}}}{\sum}\left(P_{n}^{\textrm{max}}\right)^{\beta}\times\nonumber \\
 & \left\{ \textrm{tr}\left[\boldsymbol{E}_{n}^{N_{\textrm{T}}}\left(\underset{k\in\mathcal{K}_{\textrm{IR}}}{\sum}\bar{\boldsymbol{R}}_{k}+\boldsymbol{R}_{\textrm{S}}\right)\right]\right\} ^{1-\beta},\label{eq:PA_RF}
\end{align}
\begin{align}
\bar{P}_{\textrm{PS}}\left(\boldsymbol{R}_{\boldsymbol{f}}\right) & =\frac{P_{\textrm{PS}}^{\textrm{s}}}{\log\left(1+\varepsilon^{-1}\right)}\times\nonumber \\
 & \stackrel[n=1]{N_{\textrm{T}}N_{\textrm{RF}}}{\sum}\log\left[1+\varepsilon^{-1}\textrm{tr}\left(\boldsymbol{E}_{n}^{N_{\textrm{T}}N_{\textrm{RF}}}\boldsymbol{R}_{\boldsymbol{f}}\right)\right].\label{eq:PS_I_appro-2}
\end{align}
Note that \eqref{eq:PA_RF} and \eqref{eq:PS_I_appro-2} are both
concave functions of $\boldsymbol{R}_{\boldsymbol{f}}$.

Similarly, the SINR constraints in \eqref{eq:P1_Communication_SDR}
are equivalent to
\begin{align}
\textrm{tr}\left(\boldsymbol{H}_{k}\bar{\boldsymbol{R}}_{k}\right)-\Gamma_{\textrm{IR}}\underset{i\in\mathcal{K}_{\textrm{IR}},i\neq k}{\sum}\textrm{tr}\left(\boldsymbol{H}_{k}\bar{\boldsymbol{R}}_{i}\right)\nonumber \\
-\Gamma_{\textrm{IR}}\textrm{tr}\left(\boldsymbol{H}_{k}\boldsymbol{R}_{\textrm{S}}\right)\geq\Gamma_{k}^{\textrm{IR}}\sigma_{k}^{2},\forall k\in\mathcal{K}_{\textrm{IR}},\label{eq:P1_Communication_SDR_Analog}
\end{align}
which are convex w.r.t. $\left\{ \bar{\boldsymbol{R}}_{k}\right\} $
and $\boldsymbol{R}_{\textrm{S}}$. It is observed that the sensing
constraint in \eqref{eq:P1_Sensing_PreProcess} is affine to $\boldsymbol{R}_{\boldsymbol{f}}$.
The powering constraints in \eqref{eq:P1_Powering} are equivalent
to
\begin{equation}
\boldsymbol{d}_{j}^{H}\left(\underset{k\in\mathcal{K}_{\textrm{IR}}}{\sum}\bar{\boldsymbol{R}}_{k}+\boldsymbol{R}_{\textrm{S}}\right)\boldsymbol{d}_{j}\geq\tilde{\Gamma}_{j}^{\textrm{EH}},\forall j\in\mathcal{K}_{\textrm{ER}}.\label{eq:P1_Powering_PreProcess-1}
\end{equation}
 Moreover, the constraints in \eqref{eq:P1_Per_Antenna_PreProcess}
are reformulated into
\begin{equation}
\textrm{tr}\left[\boldsymbol{E}_{n}^{N_{\textrm{T}}}\left(\underset{k\in\mathcal{K}_{\textrm{IR}}}{\sum}\bar{\boldsymbol{R}}_{k}+\boldsymbol{R}_{\textrm{S}}\right)\right]\leq P_{n}^{\textrm{max}},\forall n\in\mathcal{N}_{\textrm{T}},\label{eq:P1_Per_Antenna_PreProcess_Analog}
\end{equation}
which are also convex w.r.t. $\left\{ \bar{\boldsymbol{R}}_{k}\right\} $
and $\boldsymbol{R}_{\textrm{S}}$. Finally, we relax the constant
modulus constraints in \eqref{eq:P1_ConstantModulus} as
\begin{equation}
0\leq\left[\boldsymbol{R}_{\boldsymbol{f}}\right]_{n,n}\leq\frac{1}{N_{\textrm{T}}},\forall n\in\left\{ 1,\ldots,N_{\textrm{T}}N_{\textrm{RF}}\right\} ,\label{eq:Analog_Constant_Modulus}
\end{equation}
which are convex intervals. Thanks to the above transformation, both
the objective function and the constraints are restated as functions
of $\boldsymbol{R}_{\boldsymbol{f}}$. As a result, problem (P4) is
approximated to problem (SDR4) in the following by relaxing the rank-one
constraint $\mathrm{rank}\left(\boldsymbol{R}_{\boldsymbol{f}}\right)=1$,
in which $\boldsymbol{R}_{\boldsymbol{f}}$ is optimized instead of
$\boldsymbol{F}$.
\begin{eqnarray*}
\left(\mathrm{SDR4}\right): & \underset{\left\{ t_{i}\right\} ,\boldsymbol{R}_{\boldsymbol{f}}\succeq0}{\min} & \bar{P}_{\textrm{PA}}\left(\boldsymbol{R}_{\boldsymbol{f}}\right)+\hat{P}_{\textrm{RF}}+\bar{P}_{\textrm{PS}}\left(\boldsymbol{R}_{\boldsymbol{f}}\right)\\
 & \mathrm{s.t.} & \eqref{eq:P1_Sensing_PreProcess},\eqref{eq:P1_Communication_SDR_Analog}-\eqref{eq:Analog_Constant_Modulus}.
\end{eqnarray*}

Problem (SDR4) is still a non-convex problem due to the PA and PS
power consumption terms $\bar{P}_{\textrm{PA}}\left(\boldsymbol{R}_{\boldsymbol{f}}\right)$
and $\bar{P}_{\textrm{PS}}\left(\boldsymbol{R}_{\boldsymbol{f}}\right)$
in the objective function which are concave w.r.t. $\boldsymbol{R}_{\boldsymbol{f}}$.
In the following, we appply the SCA technique to address (SDR4). For
a specific SCA iteration $j\geq1$, let $\boldsymbol{R}_{\boldsymbol{f}}^{\left(j\right)}$
denote the local point in the $j$-th SCA iteration. First, we approximate
the PA and PS power consumption terms, $\bar{P}_{\textrm{PA}}\left(\boldsymbol{R}_{\boldsymbol{f}}\right)$
and $\bar{P}_{\textrm{PS}}\left(\boldsymbol{R}_{\boldsymbol{f}}\right)$,
as their upper bounds $\tilde{P}_{\textrm{PA}}^{\left(j\right)}\left(\boldsymbol{R}_{\boldsymbol{f}}\right)$
in \eqref{eq:SCA_PA_F} and $\tilde{P}_{\textrm{PS}}^{\left(j\right)}\left(\boldsymbol{R}_{\boldsymbol{f}}\right)$
in \eqref{eq:SCA_PS}, respectively, as shown at the top of next page.
In each SCA iteration $j$, we aim to solve the following problem.

\begin{figure*}[tbh]
\begin{align}
\tilde{P}_{\textrm{PA}}^{\left(j\right)}\left(\boldsymbol{R}_{\boldsymbol{f}}\right) & =\underset{n\in\mathcal{N_{\textrm{T}}}}{\sum}\frac{\left(P_{n}^{\textrm{max}}\right)^{\beta}}{\eta^{\textrm{max}}}\left\{ \textrm{tr}\left[\boldsymbol{E}_{n}^{N_{\textrm{T}}}\left(\mathop{\underset{k\in\mathcal{K}_{\textrm{IR}}}{\sum}}\bar{\boldsymbol{R}}_{k}^{\left(j\right)}+\boldsymbol{R}_{\textrm{S}}^{\left(j\right)}\right)\right]\right\} ^{1-\beta}+\left(1-\beta\right)\underset{n\in\mathcal{N}_{\textrm{T}}}{\sum}\frac{\left(P_{n}^{\textrm{max}}\right)^{\beta}}{\eta^{\textrm{max}}}\times\nonumber \\
 & \left\{ \textrm{tr}\left[\boldsymbol{E}_{n}^{N_{\textrm{T}}}\left(\mathop{\underset{k\in\mathcal{K}_{\textrm{IR}}}{\sum}}\bar{\boldsymbol{R}}_{k}^{\left(j\right)}+\boldsymbol{R}_{\textrm{S}}^{\left(j\right)}\right)\right]\right\} ^{-\beta}\textrm{tr}\left\{ \boldsymbol{E}_{n}^{N_{\textrm{T}}}\left[\underset{k\in\mathcal{K}_{\textrm{IR}}}{\sum}\left(\bar{\boldsymbol{R}}_{k}-\bar{\boldsymbol{R}}_{k}^{\left(j\right)}\right)+\left(\boldsymbol{R}_{\textrm{S}}-\boldsymbol{R}_{\textrm{S}}^{\left(j\right)}\right)\right]\right\} \label{eq:SCA_PA_F}
\end{align}

\begin{multline}
\tilde{P}_{\textrm{PS}}^{\left(j\right)}\left(\boldsymbol{R}_{\boldsymbol{f}}\right)=\frac{P_{\textrm{PS}}^{\textrm{s}}}{\log\left(1+\varepsilon^{-1}\right)}\stackrel[n=1]{N_{\textrm{T}}N_{\textrm{RF}}}{\sum}\left\{ \log\left[1+\varepsilon^{-1}\textrm{tr}\left(\boldsymbol{E}_{n}^{N_{\textrm{T}}N_{\textrm{RF}}}\boldsymbol{R}_{\boldsymbol{f}}^{\left(j\right)}\right)\right]+\frac{1}{\textrm{tr}\left(\boldsymbol{E}_{n}^{N_{\textrm{T}}N_{\textrm{RF}}}\boldsymbol{R}_{\boldsymbol{f}}^{\left(j\right)}\right)+\varepsilon}\textrm{tr}\left[\boldsymbol{E}_{n}^{N_{\textrm{T}}N_{\textrm{RF}}}\left(\boldsymbol{R}_{\boldsymbol{f}}-\boldsymbol{R}_{\boldsymbol{f}}^{\left(j\right)}\right)\right]\right\} \label{eq:SCA_PS}
\end{multline}

\hrule
\end{figure*}

\begin{eqnarray*}
\left(\mathrm{SCA4}.\ensuremath{j}\right): & \underset{\left\{ t_{i}\right\} ,\boldsymbol{R}_{\boldsymbol{f}}\succeq0}{\min} & \tilde{P}_{\textrm{PA}}^{\left(j\right)}\left(\boldsymbol{R}_{\boldsymbol{f}}\right)+P_{\textrm{RF}}+\tilde{P}_{\textrm{PS}}^{\left(j\right)}\left(\boldsymbol{R}_{\boldsymbol{f}}\right)\\
 & \mathrm{s.t.} & \eqref{eq:P1_Sensing_PreProcess},\eqref{eq:P1_Communication_SDR_Analog}-\eqref{eq:Analog_Constant_Modulus}.
\end{eqnarray*}
Problem (SCA4.\textit{j}) is a convex problem and can be efficiently
solved via standard convex optimization tools like CVX. Let $\boldsymbol{R}_{\boldsymbol{f}}^{*\left(j\right)}$
denote the optimal solution to problem (SCA4.\textit{j}) which is
used as the local point in the next iteration. Note that $\tilde{P}_{\textrm{PA}}^{\left(j\right)}\left(\boldsymbol{R}_{\boldsymbol{f}}\right)$
in \eqref{eq:SCA_PA_F} and $\tilde{P}_{\textrm{PS}}^{\left(j\right)}\left(\boldsymbol{R}_{\boldsymbol{f}}\right)$
in \eqref{eq:SCA_PS} serve as the upper bounds for $\bar{P}_{\textrm{PA}}\left(\boldsymbol{R}_{\boldsymbol{f}}\right)$
in \eqref{eq:PA_RF} and $\bar{P}_{\textrm{PS}}\left(\boldsymbol{R}_{\boldsymbol{f}}\right)$
in \eqref{eq:PS_I_appro-2}, respectively, which imply that the optimal
objective value for problem (SCA4.\textit{j}) in the $\left(j+1\right)$-th
iteration is consistently less than or equal to that in the $j$-th
iteration. As a result, the power consumption value for problem (SDR4)
diminishes monotonically with the iteration of SCA, thus ensuring
the convergence of the proposed algorithm. Let $\boldsymbol{\bar{R}}_{\boldsymbol{f}}^{*}$
denote the converged solution for the analog beamformer covariance,
typically of high rank. 

Finally, we construct an efficient rank-one solution to problem (P4)
by using Gaussian randomization. Specifically, we generate several
random realizations $\bar{\boldsymbol{f}^{*}}\sim\mathcal{CN}\left(\boldsymbol{0},\boldsymbol{\bar{R}}_{\boldsymbol{f}}^{*}\right)$,
constructing a set of candidate constant modulus solutions as

\begin{equation}
\boldsymbol{f}^{*}=\frac{1}{\sqrt{N_{\textrm{T}}}}e^{j\textrm{arg}\left(\bar{\boldsymbol{f}^{*}}\right)}.
\end{equation}
Subsequently, we find the optimal $\boldsymbol{f}^{*}$ among all
generated $\boldsymbol{f}^{*}$'s that yields the minimum objective
value for problem (P4) while guaranteeing the constraints in \eqref{eq:P1_Communication}-\eqref{eq:P1_PerAntennaPower}.
The optimized solution $\boldsymbol{F}^{*}$ is then recovered by
reshaping the optimal $\boldsymbol{f}^{*}$ into the form of an analog
beamformer matrix. We then use the corresponding $\bar{\boldsymbol{f}^{*}}$
of the optimal $\boldsymbol{f}^{*}$ to design the on-off control
for PSs in the next subsection.

\subsection{On-off Control based on Beamforming Weights}

In this subsection, we propose an efficient method based on beamforming
weights to determine the on-off control of RF chains and PSs by using
the obtained beamforming solution $\left\{ \boldsymbol{w}_{k}^{*}\right\} $,
$\boldsymbol{S}^{*}$, and $\boldsymbol{F}^{*}$. Heuristically, we
prioritize switching off components with minimum impact on the system
performance. Specifically, we initially attempt to switch off as many
RF chains as possible, and then select PSs to further minimize the
total power consumption.

First, we consider the on-off control of RF chains based on their
beamforming weights. Specifically, we define $v_{n}=\sum_{k\in\mathcal{K}_{\textrm{IR}}}\left|\boldsymbol{1}_{n}^{N_{\textrm{RF}}}\boldsymbol{w}_{k}\right|^{2}+\textrm{tr}\left(\boldsymbol{E}_{n}^{N_{\textrm{RF}}}\boldsymbol{S}\right)$
as the beamforming weight for each RF chain $n\in\mathcal{N}_{\textrm{RF}}$.
The RF chain with a smaller $v_{n}$ is switched off with priority.
In a detailed procedure, we sort the beamforming weights of all RF
chains and sequentially switch off the corresponding RF chains in
ascending order. Following the re-implementation of the beamforming
optimization, we compare the resulting power consumption values. Subsequently,
we select the RF chain on-off configuration that achieves the lowest
overall power consumption, thereby designating it as the optimized
configuration.

Then, we implement the on-off control of PSs based on their beamforming
weights. We consider the optimal $\boldsymbol{f}^{*}$ and its corresponding
$\bar{\boldsymbol{f}^{*}}$. By reshaping $\bar{\boldsymbol{f}^{*}}$
into its matrix form $\bar{\boldsymbol{F}^{*}}$, we define $c_{i,j}=\left|\left[\bar{\boldsymbol{F}^{*}}\right]_{i,j}\right|$
as the beamforming weight for each $(i,j)$-th PS element. The PS
with a smaller $c_{i,j}$ takes priority to be switched off. Specifically,
we sort the beamforming weights of all PS elements and sequentially
switch off the corresponding PS elements in ascending order. Subsequently,
we compare the resultant power consumption values after re-implementing
the beamforming optimization, and then select the PS on-off configuration
that yields the lowest overall power consumption, thus designating
it as the optimized configuration. During the re-implementation of
beamforming, the analog beamforming process is re-executed by considering
the additional constraint on the $n$-th deactivated PS element $\left[\boldsymbol{R}_{\boldsymbol{f}}\right]_{n,n}=0$.
After implementing the above RF chains and PSs on-off control procedures,
we can obtain a highly energy-efficient on-off switching design for
the whole system of RF chains and PSs.

\section{Numerical Results}

In this section, numerical results are provided to verify the performance
of our proposed design. In the simulations, we set the numbers of
antennas at the transmitter, antennas at the sensing receiver, and
RF chains at the transmitter as $N_{\textrm{T}}=32$, $N_{\textrm{S}}=32$,
and $N_{\textrm{RF}}=16$, respectively. We also set the number of
IRs, sensing targets, and ERs as $K_{\textrm{IR}}=6$, $K_{\textrm{S}}=5$,
and $K_{\textrm{ER}}=5$, respectively. Moreover, the noise power
at each IR and the sensing receiver are set to be $\sigma_{k}^{2}=\sigma_{\textrm{S}}^{2}=-103\,\textrm{dBm},k\in\mathcal{K}_{\textrm{IR}}$.
The power consumption of a single RF chain and a single PS are set
as $P_{\textrm{RF}}^{\textrm{s}}=0.5\,\textrm{W}$ and $P_{\textrm{PS}}^{\textrm{s}}=42\,\textrm{mW}$,
respectively, with the static power consumption by other components
set as $P_{\textrm{static}}=10\,\textrm{W}$. We set the maximum transmit
power at each antenna $n\in\mathcal{N}_{\textrm{T}}$ as $P_{n}^{\textrm{max}}=1.5\,\textrm{W}$,
the maximum PA efficiency as $\eta^{\textrm{max}}=0.38$, and the
efficiency factor as $\beta=0.5$, respectively. We assume the SINR
constraints at different users and the minimum harvested power level
at different ERs are identical. We adopt the Rayleigh fading channel
model for wireless channel between the BS and IRs, the line-of-sight
(LoS) channel model for the channel between the BS and sensing targets,
and the Rician fading channel model with Rician factor being $3\,\textrm{dB}$
for the wireless channel between the BS and ERs. The large-scale path
loss is modelled as $51.2+41.2\log_{10}r$, with $r$ in meters denoting
the distance. If not specified, the distances from the BS transmitter
to the IRs, sensing targets, and ERs are set as $50\,\textrm{m}$,
$50\,\textrm{m}$, and $10\,\textrm{m}$, respectively. The target
angles $\left\{ \theta_{i}\right\} _{i=1}^{K_{\textrm{S}}}$ are generated
randomly. The transmission frame length is set as $L=30$. In the
non-linear EH model, we set $M_{j}=0.02$, $a_{j}=6400$, and $b_{j}=0.003$,
determined through curve fitting based on measurement data \cite{4494663}.

We consider the following benchmark schemes in our evaluation, comparing
them with our proposed \textbf{joint hybrid beamforming (BF) and on-off
control} design.
\begin{itemize}
\item \textbf{Hybrid BF w/o on-off control}: This corresponds to the case
when the dynamic on-off control is totally ignored and all RF chains
and PSs are activated. The hybrid beamformers are optimized by the
proposed algorithm.
\item \textbf{PS on-off control only}: This corresponds to the case when
the dynamic on-off control of RF chains is ignored and actually determined
by the on-off status of the corresponding PSs.
\item \textbf{RF chain on-off control only}: This corresponds to the case
when the dynamic on-off control of PSs is ignored while the effect
of RF chain on-off switching is solely concerned.
\item \textbf{Digital BF w/ on-off control}: This corresponds to our proposed
joint design when applied in the fully-digital transmitter architecture.
In this case, the analog beamformer $\boldsymbol{F}$ and digital
beamformer $\left\{ \boldsymbol{w}_{k}\right\} $ are substituted
by a combined fully-digital beamformer, and the on-off control of
the transmitter chains, each comprising an RF chain and a PA, is considered.
The problem is solved similarly to the proposed algorithm, based on
SDR and SCA.
\item \textbf{Case w/ fixed PA efficiency}: This corresponds to our proposed
joint design by setting $\beta=0$. In this case, the PA power consumption
model is reduced to the following form:
\begin{align}
 & P_{\textrm{PA}}\left(\boldsymbol{F},\left\{ \boldsymbol{w}_{k}\right\} ,\boldsymbol{S}\right)=\frac{1}{\eta^{\textrm{max}}}\times\nonumber \\
 & \underset{n\in\mathcal{N}_{\textrm{T}}}{\sum}\left\{ \underset{k\in\mathcal{K}_{\textrm{IR}}}{\sum}\left|\boldsymbol{1}_{n}^{N_{\textrm{T}}}\boldsymbol{F}\boldsymbol{w}_{k}\right|^{2}+\textrm{tr}\left(\boldsymbol{E}_{n}^{N_{\textrm{T}}}\boldsymbol{F}\boldsymbol{S}\boldsymbol{F}^{H}\right)\right\} .
\end{align}
The optimization problem is then similarly formulated and solved by
adopting the proposed algorithm.
\end{itemize}
\begin{figure}[tbh]
\begin{centering}
\includegraphics[width=3.5in]{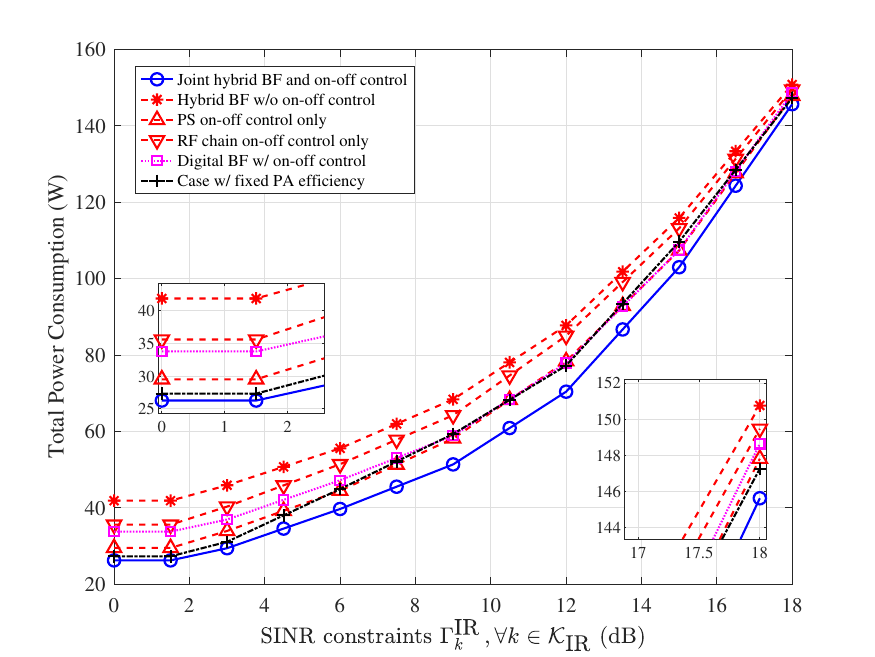}
\par\end{centering}
\caption{The total power consumption w.r.t SINR requirements with $\Gamma_{\textrm{S}}=0.1$
and $\Gamma_{j}^{\textrm{DC}}=-2\,\textrm{dBm},\forall j\in\mathcal{K}_{\textrm{ER}}$.\label{Fig. 2}}
\end{figure}

Fig. 2 shows the total power consumption concerning the SINR requirements,
given a CRB threshold $\Gamma_{\textrm{S}}=0.1$ and EH levels $\Gamma_{j}^{\textrm{DC}}=-2\,\textrm{dBm},\forall j\in\mathcal{K}_{\textrm{ER}}$.
It is observed that the power consumption of our proposed joint design
is markedly lower than that of other benchmark schemes. The dynamic
on-off control of RF chains and PSs is shown to be particularly beneficial
for saving power consumption when the SINR requirements become less
stringent. This is due to the fact that more RF chains and PSs are
necessarily active with high SINR requirements. Furthermore, the tradeoff
between communication and the other two functionalities is elucidated.
When the SINR constraints $\Gamma_{k}^{\textrm{IR}},\forall k\in\mathcal{K}_{\textrm{IR}},$
are lower than a certain turning point, the total power consumption
is primarily influenced by CRB or EH constraints. This manifests as
a horizontal trend in the total power consumption w.r.t. SINR, highlighting
the intricate balance between the various performance requirements.
Moreover, it is observed that the total power consumption by fully-digital
beamforming is higher than that by our proposed HAD architecture,
indicating that the utilization of low-power PSs in analog beamforming
can lead to a reduced number of RF chains, by partially offsetting
the demand for digital beamforming. 

\begin{figure}[tbh]
\begin{centering}
\includegraphics[width=3.5in]{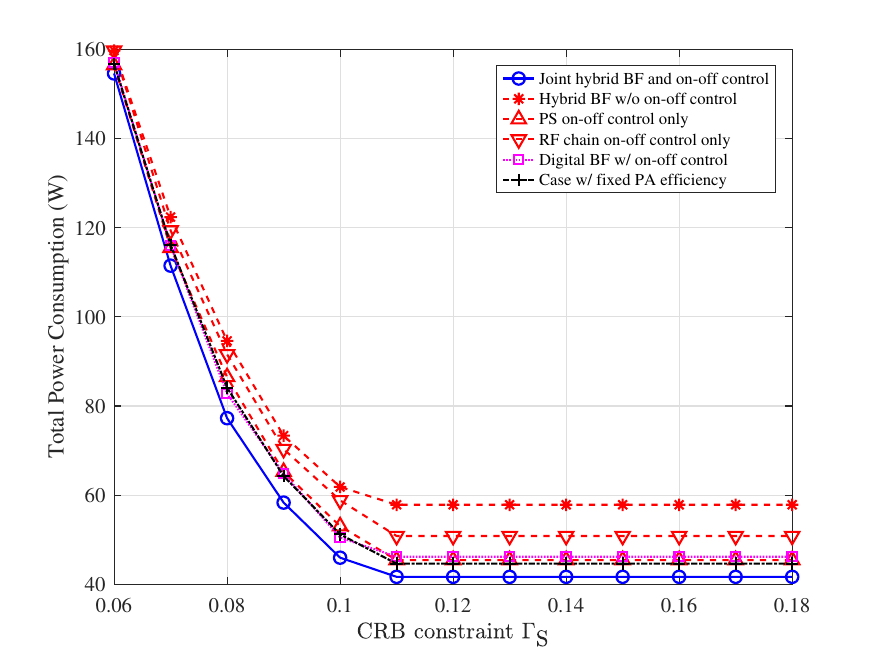}
\par\end{centering}
\caption{The total power consumption w.r.t CRB threshold with $\Gamma_{k}^{\textrm{IR}}=6\,\textrm{dB},\forall k\in\mathcal{K}_{\textrm{IR}},$
and $\Gamma_{j}^{\textrm{DC}}=-2\,\textrm{dBm},\forall j\in\mathcal{K}_{\textrm{ER}}$.\label{Fig. 3}}
\end{figure}

\begin{figure}[tbh]
\begin{centering}
\includegraphics[width=3.5in]{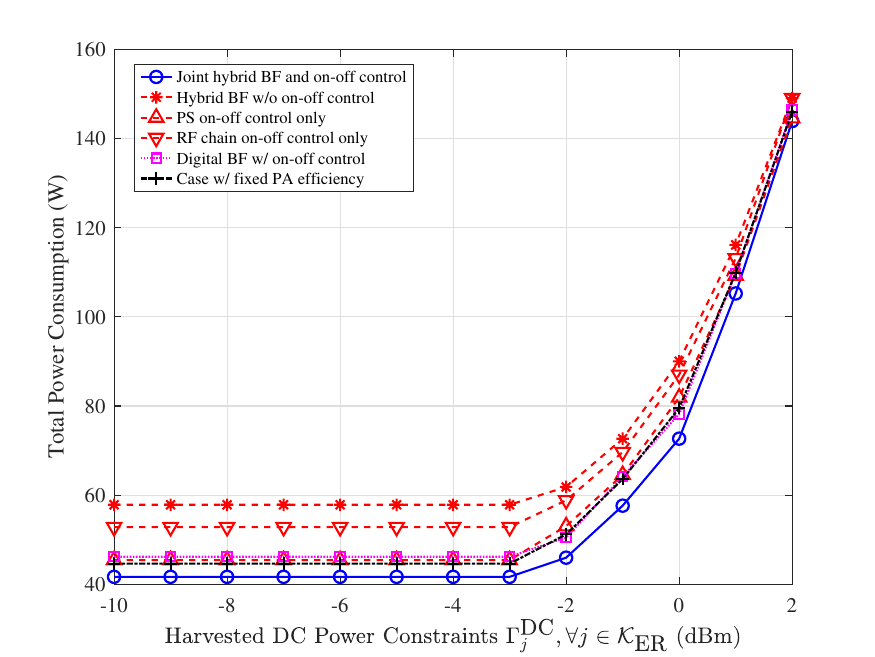}
\par\end{centering}
\caption{The total power consumption w.r.t EH constraint with $\Gamma_{k}^{\textrm{IR}}=6\,\textrm{dB},\forall k\in\mathcal{K}_{\textrm{IR}},$
and $\Gamma_{\textrm{S}}=0.1$.\label{Fig. 4}}

\end{figure}

Fig. 3 and Fig. 4 depict the total power consumption concerning
the CRB threshold under given SINR requirements $\Gamma_{k}^{\textrm{IR}}=6\,\textrm{dB},\forall k\in\mathcal{K}_{\textrm{IR}},$
and EH levels $\Gamma_{j}^{\textrm{DC}}=-2\,\textrm{dBm},\forall j\in\mathcal{K}_{\textrm{ER}}$,
and concerning the EH constraint under given SINR requirements $\Gamma_{k}^{\textrm{IR}}=6\,\textrm{dB},\forall k\in\mathcal{K}_{\textrm{IR}},$
and a CRB threshold $\Gamma_{\textrm{S}}=0.1$, respectively. Our
proposed design consistently outperforms all benchmark schemes in
terms of power-saving performance. It is observed that when the CRB
threshold is high or the EH constraints are low, the total power consumption
is dominated by the other two requirements. By combining such phenomenon
in Fig. 2, Fig. 3, and Fig. 4, we explicitly observe the tradeoff
among the three functionalities, including communication, sensing,
and powering.

\begin{figure}[tbh]
\begin{centering}
\includegraphics[width=3.5in]{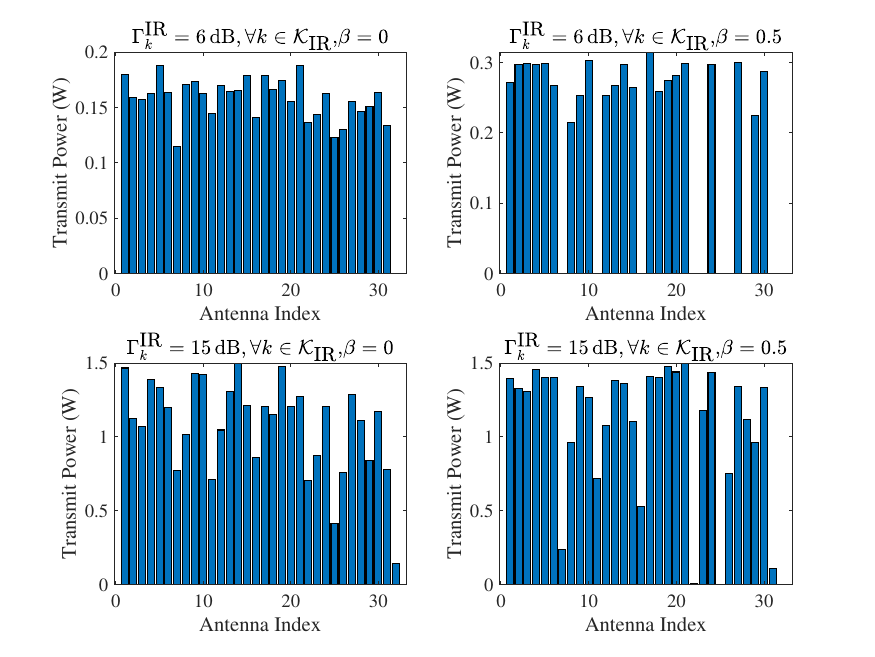}
\par\end{centering}
\caption{The optimized transmit power allocation across various antennas, with
$\Gamma_{\textrm{S}}=0.08$ and $\Gamma_{j}^{\textrm{DC}}=0\,\textrm{dBm},\forall j\in\mathcal{K}_{\textrm{ER}}$.\label{Fig. 5}}
\end{figure}

Fig. 5 shows a comparative analysis of the optimized transmit power
allocation across various antennas, considering non-linear PA efficiency
(i.e., $\beta=0.5$) versus fixed PA efficiency (i.e., $\beta=0$).
In the traditional model employing fixed PA efficiency, the prevalent
practice is to activate almost all antennas while judiciously distributing
transmit power. On the contrary, our proposed design unravels a distinctive
pattern, where certain antennas are deactivated with zero transmit
power, while others receive augmented power allocation to meet the
performance constraints. This phenomenon arises from the non-linear
nature of PA efficiency, which amplifies in tandem with the output
signal power at each antenna. Consequently, the BS exhibits a tendency
to deactivate more antennas and elevate the transmit power on the
remaining antennas, aiming to exploit heightened PA efficiency. Moreover,
comparing the cases with $\Gamma_{k}^{\textrm{IR}}=6\,\textrm{dB},\forall k\in\mathcal{K}_{\textrm{IR}},$
and $\Gamma_{k}^{\textrm{IR}}=15\,\textrm{dB},\forall k\in\mathcal{K}_{\textrm{IR}}$,
we further observe that the distribution of power allocation at different
transmit antennas turns out to be more polarized with a higher performance
requirement.

\begin{figure}[tbh]
\begin{centering}
\includegraphics[width=3.5in]{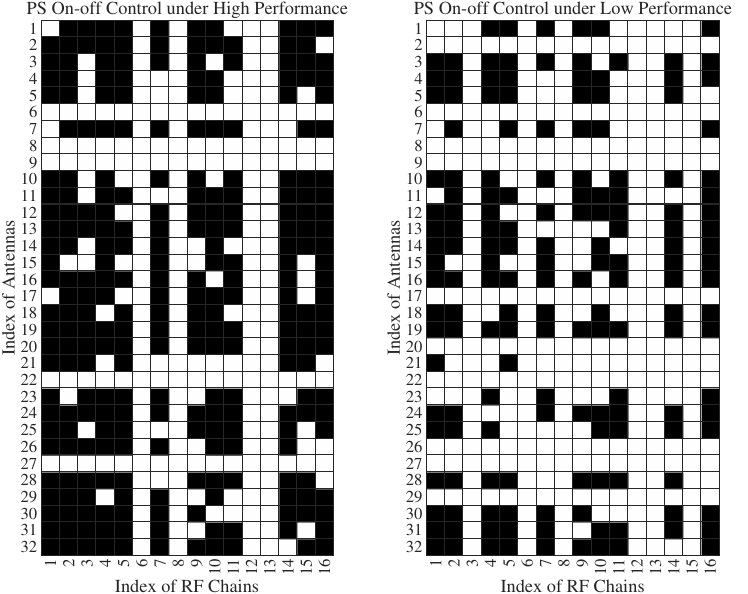}
\par\end{centering}
\caption{The on-off status of PSs, with black denoting \textquotedblleft on\textquotedblright{}
and white denoting \textquotedblleft off\textquotedblright . The left
subfigure is with $\Gamma_{k}^{\textrm{IR}}=12\,\textrm{dB},\forall k\in\mathcal{K}_{\textrm{IR}}$,
$\Gamma_{\textrm{S}}=0.08$, and $\Gamma_{j}^{\textrm{DC}}=0\,\textrm{dBm},\forall j\in\mathcal{K}_{\textrm{ER}}$.
The right subfigure is with $\Gamma_{k}^{\textrm{IR}}=6\,\textrm{dB},\forall k\in\mathcal{K}_{\textrm{IR}}$,
$\Gamma_{\textrm{S}}=0.1$, and $\Gamma_{j}^{\textrm{DC}}=-2\,\textrm{dBm},\forall j\in\mathcal{K}_{\textrm{ER}}$.\label{Fig. 6}}
\end{figure}

Fig. 6 shows the on-off control phenomenon on the PSs via an intuitive
image, where black blocks are ``on'' PSs with unit norm and white
blocks are ``off'' PSs being zero. It is observed that our proposed
HAD architecture and on-off control algorithm efficiently switches
off the PSs to save power. Some antennas are actually switched off
due to the on-off control of PSs, while the others are allocated concentrated
transmit power to exploit the non-linear PA efficiency. This is due
to the fact that, induced by the non-linear PA efficiency, the corresponding
PSs are with low beamforming weights and thus preferentially switched
off. By comparing the two subfigures under high and low performance
constraints, respectively, we notice that the PSs, which are deactivated
when the requirements are high, are also switched off when the requirements
are less stringent. In the left subfigure, we turn off $4$ RF chains
and $248$ PSs in total to save power, where $5$ antennas are actually
switched off, while in the right subfigure, an additional $2$ RF
chains and $101$ PSs are switched off, contributing to a further
reduced total power consumption. The PSs with the least beamforming
weights are switched off to save more power, while other PSs remain
active to guarantee the performance.

\section{Conclusion}

In this paper, we studied the energy-efficient hybrid beamforming
design for a multi-functional ISCAP system, where a BS sends unified
wireless signals to communicate with multiple IRs, sense multiple
point targets, and wirelessly charge multiple ERs concurrently. We
presented a novel HAD architecture with a switch network for the BS
transmitter to enable the dynamic on-off control of its RF chains
and PSs. We introduced a practical and comprehensive power consumption
model for the BS, accounting for the power-dependent non-linear PA
efficiency and the on-off non-transmission power consumption model
of RF chains and PSs. We proposed an efficient joint hybrid beamforming
and dynamic on-off control algorithm to minimize the total power consumption
of the BS, while guaranteeing the performance constraints on communication
rates, sensing CRB, and harvested power levels, as well as the per-antenna
transmit power constraint and the constant modulus constraints for
the analog beamformer, by employing AO, SCA, and SDR techniques. Then,
based on the optimized beamforming weights, we developed an efficient
method to determine the on-off control of RF chains and PSs, and update
the associated hybrid beamforming solution. Numerical results showed
that the proposed design significantly improves the energy efficiency
for ISCAP compared to other benchmark schemes without joint design
of hybrid beamforming and dynamic on-off control. A tradeoff among
the three functionalities was revealed, and the benefit of dynamic
on-off control in energy reduction was validated, especially under
loose multi-functional performance requirements.

\appendices{}

\section{Proof of Proposition 1}

It follows from \eqref{eq:Rank-1-S} that $\sum_{k\in\mathcal{K}_{\textrm{IR}}}\boldsymbol{R}_{k}^{*}+\boldsymbol{S}^{*}=\sum_{k\in\mathcal{K}_{\textrm{IR}}}\boldsymbol{\bar{R}}_{k}^{*}+\boldsymbol{\bar{S}}^{*}$,
and as a result, $\left\{ \boldsymbol{R}_{k}^{*}\right\} $, $\boldsymbol{S}^{*}$
and $\left\{ \boldsymbol{\bar{R}}_{k}^{*}\right\} $, $\boldsymbol{\bar{S}}^{*}$
achieve the same objective values and both satisfy the constraints
in \eqref{eq:P1_Sensing_PreProcess}-\eqref{eq:P1_Per_Antenna_PreProcess}.
Next, note that the SINR constraints in \eqref{eq:P1_Communication_SDR}
are equivalent to
\begin{equation}
\frac{\textrm{tr}\left(\boldsymbol{H}_{k}\boldsymbol{F}\boldsymbol{R}_{k}\boldsymbol{F}^{H}\right)}{\textrm{tr}\left[\boldsymbol{H}_{k}\boldsymbol{F}\left(\underset{i\in\mathcal{K}_{\textrm{IR}}}{\sum}\boldsymbol{R}_{i}+\boldsymbol{S}\right)\boldsymbol{F}^{H}\right]+\sigma_{\textrm{IR}}^{2}}\geq\frac{\Gamma_{\textrm{IR}}}{1+\Gamma_{\textrm{IR}}},\forall k\in\mathcal{K}_{\textrm{IR}}.\label{eq:SINR_Proposition_SDR}
\end{equation}
It is verified from \eqref{eq:Rank-1-wk} that
\begin{equation}
\textrm{tr}\left(\boldsymbol{H}_{k}\boldsymbol{F}\boldsymbol{R}_{k}^{*}\boldsymbol{F}^{H}\right)=\textrm{tr}\left(\boldsymbol{H}_{k}\boldsymbol{F}\boldsymbol{\bar{R}}_{k}^{*}\boldsymbol{F}^{H}\right).
\end{equation}
Therefore, $\left\{ \boldsymbol{R}_{k}^{*}\right\} $ also satisfies
the SINR constraints in \eqref{eq:SINR_Proposition_SDR}. Furthermore,
for any $\boldsymbol{v}\in\mathbb{C}^{N_{\textrm{T}}\times1}$, it
holds that
\begin{align}
 & \boldsymbol{v}^{H}\boldsymbol{F}\left(\boldsymbol{\bar{R}}_{k}^{*}-\boldsymbol{R}_{k}^{*}\right)\boldsymbol{F}^{H}\boldsymbol{v}\nonumber \\
 & =\boldsymbol{v}^{H}\boldsymbol{F}\boldsymbol{\bar{R}}_{k}^{*}\boldsymbol{F}^{H}\boldsymbol{v}-\left|\boldsymbol{v}^{H}\boldsymbol{F}\boldsymbol{\bar{R}}_{k}^{*}\boldsymbol{F}^{H}\boldsymbol{h}_{k}\right|^{2}\left(\boldsymbol{h}_{k}^{H}\boldsymbol{F}\boldsymbol{\bar{R}}_{k}^{*}\boldsymbol{F}^{H}\boldsymbol{h}_{k}\right)^{-1}.
\end{align}
According to the Cauchy-Schwarz inequality, we have
\begin{equation}
\left(\boldsymbol{v}^{H}\boldsymbol{F}\boldsymbol{\bar{R}}_{k}^{*}\boldsymbol{F}^{H}\boldsymbol{v}\right)\left(\boldsymbol{h}_{k}^{H}\boldsymbol{F}\boldsymbol{\bar{R}}_{k}^{*}\boldsymbol{F}^{H}\boldsymbol{h}_{k}\right)\geq\left|\boldsymbol{v}^{H}\boldsymbol{F}\boldsymbol{\bar{R}}_{k}^{*}\boldsymbol{F}^{H}\boldsymbol{h}_{k}\right|^{2},
\end{equation}
and it follows that $\boldsymbol{v}^{H}\boldsymbol{F}\left(\boldsymbol{\bar{R}}_{k}^{*}-\boldsymbol{R}_{k}^{*}\right)\boldsymbol{F}^{H}\boldsymbol{v}\geq0$.
Accordingly, we have $\boldsymbol{\bar{R}}_{k}^{*}-\boldsymbol{R}_{k}^{*}\succeq0$.
Additionally, the positive semidefinite constraint holds for $\left\{ \boldsymbol{R}_{k}^{*}\right\} $
and $\boldsymbol{S}^{*}$ since the summation of a set of positive
semidefinite matrices is positive semidefinite, implying $\boldsymbol{S}^{*}\succeq0$.
Notice that $\mathrm{rank}\left(\boldsymbol{R}_{k}^{*}\right)\leq1$
with $\boldsymbol{R}_{k}^{*}=\boldsymbol{w}_{k}^{*}\left(\boldsymbol{w}_{k}^{*}\right)^{H}$.
Therefore, $\left\{ \boldsymbol{R}_{k}^{*}\right\} $ and $\boldsymbol{S}^{*}$
are optimal for problem (SDR2). As a result, Proposition 1 is proved.

\bibliographystyle{IEEEtran}
\bibliography{IEEEabrv,IEEEexample,reference}

\end{document}